\documentclass[twocolumn,amsmath,amssymb,10pt,a4paper,superscriptaddress,nofootinbib]{revtex4}
\pdfoutput=1
\usepackage{geometry}
\usepackage{graphicx}
\usepackage{dcolumn}
\usepackage{bm}
\usepackage{enumerate}

\newcommand{\be}{\begin{equation}}
\newcommand{\ee}{\end{equation}}
\newcommand{\bd}{\begin{displaymath}}
\newcommand{\ed}{\end{displaymath}}
\newcommand{\BE}{\begin{eqnarray}}
\newcommand{\EE}{\end{eqnarray}}

\newcommand{\bx}{\ensuremath{\mathbf{x}}}

\newcommand{\bn}{\ensuremath{\mathbf{n}}}

\newcommand{\boldeta}{{\mbox{\boldmath $\eta$}}}

\newcommand{\bxi}{\bm{\xi}}

\newcommand{\avg}[1]{\left\langle{#1}\right\rangle}

\begin{document}

\title{Gaussian approximations for stochastic systems with delay:\\ chemical Langevin equation and application to a Brusselator system}

\author {Tobias Brett}
\email{tobias.brett@postgrad.manchester.ac.uk }
\affiliation{Theoretical Physics, School of Physics and Astronomy, The University of Manchester, Manchester M13 9PL, United Kingdom}

\author{Tobias Galla}
\email{tobias.galla@manchester.ac.uk}
\affiliation{Theoretical Physics, School of Physics and Astronomy, The University of Manchester, Manchester M13 9PL, United Kingdom}
 
\begin{abstract}
We present a heuristic derivation of Gaussian approximations for stochastic chemical reaction systems with distributed delay. In particular we derive the corresponding chemical Langevin equation. Due to the non-Markovian character of the underlying dynamics these equations are integro-differential equations, and the noise in the Gaussian approximation is coloured. Following on from the chemical Langevin equation a further reduction leads to the linear-noise approximation. We apply the formalism to a delay variant of the celebrated Brusselator model, and show how it can be used to characterise noise-driven quasi-cycles, as well as noise-triggered spiking. We find surprisingly intricate dependence of the typical frequency of quasi-cycles on the delay period. \end{abstract}

\maketitle

\section{Introduction}
 
Traditionally, chemical reaction systems are modelled by sets of differential equations, also known as rate equations. These equations describe the time evolution of the  continuous real-valued concentrations of the different particle types, and they are derived from the microscopic reactions using mass-action principles \cite{benson}. This framework is mathematically convenient: the theory of ordinary and partial differential equations is well developed, and the tools for their analysis are readily available. Descriptions based on deterministic differential equations have one major drawback though, they systematically neglect all effects of stochasticity. Mathematically, deterministic descriptions are only appropriate for large (formally infinite) systems.

It is now commonly accepted that intrinsic stochasticity, arising due to the finite numbers of particles in reaction systems, can have significant effects on the dynamics \cite{intrinsicnoise}. This includes noise-induced cycles \cite{newman_mckane}, patterns and waves \cite{butler,biancalani}, and phenomena such as extinction and fixation \cite{nowak}. In order to fully capture these effects models must describe the reaction dynamics at the micro-scale and keep track of integer-valued particle numbers. Assuming a well-mixed reactor, the state of the system is fully characterised by the number of molecules of each type present in the system at a given time. The time dependence of the probability distribution over the space of states is then governed by a chemical master equation (CME) \cite{nordsieck, vankampen}.

The CME is in general difficult to solve, exact solutions only exist for a limited set of examples, such as simple one-step birth-death processes and those satisfying detailed balance \cite{vankampen, gardiner}. For the majority of cases approximative schemes present the best opportunity for analytical progress. Deterministic rate equations, as described above, are  the simplest such approximation. They neglect all stochasticity, and formulate an effective dynamics for the first moment of the probability distribution over microstates. These equations can either be written down based on intuition, or they can formally be derived from the CME by means of the so-called system-size expansion. This requires the presence of a large parameter, typically the volume of the reactor  or the total particle number. Its inverse is then a small quantity, and serves as an expansion parameter. The deterministic approximation is obtained from the lowest order of this expansion, corresponding to an infinite volume or particle number \cite{vankampen}. Retaining the sub-leading terms in the expansion on the other hand ultimately leads to a stochastic differential equation with Gaussian noise. There are multiple methods by which the expansion of the CME can be carried out, e.g. the method by van Kampen \cite{vankampen} or the Kramers-Moyal expansion \cite{km}. Depending on the details one obtains either additive noise, this is referred to as the linear-noise approximation (LNA), or multiplicative noise. The resulting stochastic differential equation in the latter case is frequently referred to as the chemical Langevin equation (CLE). We will refer to any of these methods as Gaussian approximations, as the noise in the resulting effective dynamics is Gaussian, but it should be noted that the distribution of the quantities described by those effective dynamics itself may not be Gaussian in the case of multiplicative noise. The various different approximations are all closely related, see e.g. \cite{grima.2011} for further details.

The starting point for most of the expansion techniques mentioned above is the CME, which can generally only be formulated for Markovian systems. These are systems without memory, i.e. systems for which the dynamics at any one time only depends on the state of the system at that time, but not on the previous history by which the system has arrived at this state \cite{gardiner}. This implies that all effects of chemical reactions must occur instantaneously, no reaction event has any further effects on the system at later times.  The times between reaction events follow exponential statistics, or equivalently the number of events occurring in a fixed time interval is Poissonian. These assumptions do not hold in a variety of applications. One of these is gene expression dynamics, in which there is a characteristic time delay associated with transcription and translation events. These may be triggered at a given time, but their products are only generated at a later time \cite{ hirata, takashima}. Another example is in the modelling of epidemics, where an individual may be infected at a given time, and where recovery is at a later time, drawn from a distribution peaked around a typical infectious period. Recovery is then not an exponential process. We will refer to models of this type as delay models. The delay description can be considered an effective description at a coarse-grained level. On a much finer level a Markovian description of the underlying reactions may be appropriate. A key distinction is between models with a fixed delay and models with so-called distributed delay \cite{kuang}. In models with distributed delay the delay times are stochastic variables themselves, whenever a delay reaction is initiated a delay time is drawn from an underlying distribution, and delayed effects materialise at this later time. Fixed-delay models are the special case in which the delay kernel is a $\delta$-function.

Non-Markovian systems can in principle be mapped onto high-dimensional Markovian systems \cite{vankampen}. This procedure is not applicable to general delay kernels though, or it leads to Markovian dynamics with an infinite number of degrees of freedom. This is often unsatisfactory, and as a consequence many existing analyses of delay systems have focused on the deterministic limit \cite{gene.2003, roussel, hinch}. It is only more recently that a more systematic stochastic description of delay reactions has been attempted \cite{barrio,bratsun,galla,lafuerza,brett_galla}. This existing work has predominantly focused on extensions to the CME to account for delay reactions \cite{bratsun,galla,lafuerza}, but these equations typically do not close and are limited to specific classes of model systems. Systems with constant delay have been studied using the LNA in \cite{
bratsun,galla}. 

In a previous piece of work \cite{brett_galla} we proposed the use of a method known from condensed matter physics \cite{altland} to describe the time evolution of discrete reaction systems with distributed delays. This technique, the so-called Martin-Siggia-Rose-Janssen-de-Dominicis generating functional \cite{msr} takes a path-based view. It considers the space of all possible time courses of the system and formulates the probability for a given path to occur as a dynamic generating functional or path integral, effectively representing the Fourier transform of the probability measure in the space of all dynamic paths. This is a powerful formulation applicable to a wide class of delay systems, and it can be used to derive effective Gaussian approximations, in particular an equivalent of the CLE for delay systems. The approach involves relatively complex mathematics though, and the purpose of the present paper is to show how this machinery can be bypassed. We present a heuristic, more intuitive procedure to derive the CLE for delay systems. To demonstrate its utility this method is applied to a variant of the celebrated Brusselator system with delay. In particular we focus on effects of stochasticity in parameter regimes in which the deterministic delay system approaches a fixed point. We are able to characterise the stationary distribution of the system, and the noise-induced quasi-cycles the dynamics generate. From our analytical calculations we find a surprisingly intricate dependence of the typical frequency of the cycles on the delay. We also study choices of parameters in which the deterministic delay dynamics constitute an excitable system. In the presence of intrinsic noise stochastically triggered spikes are found, we show how these can be simulated efficiently using the CLE.

The remainder of the paper is organised as follows: In order to keep our paper self-contained and pedagogical we briefly summarise earlier work by Gillespie \cite{gillespie_chemlang} in Sec. \ref{sec:markov}. In particular we describe how the Gaussian approximation is obtained for Markovian systems, and how the corresponding CLE is derived. In Sec. \ref{sec:approxfordelay} we then carry out a similar analysis for delay systems, and derive both the delay-CLE and subsequently the corresponding LNA. These results are then applied to the specific example of the Brusselator with delay dynamics in Sec. \ref{sec:bruss}. We present a summary our work in Sec. \ref{sec:concl}, and outline possible future work. In the Appendix we briefly describe how the well-known modified next-reaction method is modified to accommodate distributed delays, and we provide further supplementary details of our analytical calculations.

\section{Markovian dynamics}
\label{sec:markov}

Before considering delay reactions it is useful to first summarise the main results for systems without delay obtained from the above expansion methods. We consider a well-mixed system composed of particles of different chemical species $X_\alpha$, $\alpha=1,\dots,S$. The corresponding particle numbers are written as $n_\alpha\in\mathbb{N}_0$. Interactions occur via a set of reactions, $i=1,\dots, R$. Each chemical reaction, $i$, is written in the form
\be
\sum_\alpha s_{i,\alpha}X_\alpha\overset{k_i}{\longrightarrow} \sum_\alpha q_{i,\alpha}X_\alpha.
\ee
The stoichiometric coefficient $s_{i,\alpha}$ represents the number of $X_\alpha$ particles entering the reaction, and $q_{i,\alpha}$ is the number of such particles exiting. It is useful to define the quantities $v_{i,\alpha} = q_{i,\alpha} - s_{i,\alpha}$, so that $v_{i,\alpha}$ indicates the change in the number of particles of type $\alpha$ when one reaction of type $i$ occurs. The above notation indicates that the rate with which reaction $i$ occurs is given by
\be
R_i(\bn/\Omega) = \Omega k_i\prod_{\alpha} \left(\frac{n_\alpha}{\Omega}\right)^{s_{i,\alpha}}\equiv \Omega r_i(\bn/\Omega)
\ee
when the system is in state $\bn=(n_1,\dots, n_S)$. These rates are scaled with an overall parameter $\Omega$, representing for example the volume of the system or a scale for the total number of particles. Each rate is of order $\Omega$, such that the number of reactions occurring per unit time scales as $\Omega$ as well, in-line with standard conventions \cite{vankampen}. The quantity $r_i(\bn/\Omega)$ can be understood as the `intensive' rate. The CME is then given by 
\be
\dot P_{\bn}(t) = \sum_{\bn'} T_{\bn'\bn}P_{\bn'} -T_{\bn\bn'} P_{\bn},
\ee
where $P_\bn(t)$ is the probability that the system is in state $\bn$ at time $t$. The quantity $T_{\bn\bn'}$ is the total transition rate from $\bn$ to $\bn'$, and similar for $T_{\bn'\bn}$. Next, it is convenient to introduce concentration variables $\bx=\bn/\Omega$. Carrying out a Kramers-Moyal expansion of the CME in powers of $\Omega^{-1}$ up to and including sub-leading terms the following CLE is found (via a Fokker-Planck equation) \cite{vankampen}
\be
\dot  x_\alpha(t)  = \sum_i v_{i,\alpha}r_i[\bx(t)] + \frac{1}{\sqrt{\Omega}}\eta_\alpha.
\label{eq:chemlang}
\ee
Here, $\eta_\alpha$ is a white Gaussian noise term with mean $\avg{\eta_\alpha(t)} = 0$ and with correlations across species given by  
\be
\avg{\eta_\alpha(t)\eta_\beta(t')} = \left(\sum_i v_{i,\alpha}v_{i,\beta}r_i[\bx(t)]\right) \delta(t-t').
\label{eq:noisecorrelator}
\ee

The above result for the CLE is derived from a controlled expansion procedure, we do not re-iterate the full details here. Alternatively, the same result can be obtained from a heuristic argument following the lines of \cite{gillespie_chemlang}, which we summarise here. As a first step one discretises time into intervals of duration $\Delta$. Assuming for the time being that reaction rates remain constant in each such interval, the number of reactions of type $i$ firing in the time interval $[t,t+\Delta)$ is an integer random variable $k_{i,t}$ drawn from a Poissonian distribution with parameter $R_i(\bx_t)\Delta$. One then has
\be
x_{\alpha, t+\Delta} = x_{\alpha,t} + \sum_{i} v_{\alpha,i}\frac{k_{i,t}}{\Omega}.
\label{eq:differenceeq}
\ee
The $\{k_{i,t}\}$ are statistically independent for different $i$ and $t$, and, given their Poissonian statistics, we have $\avg{k_{i,t}}=R_i(\bx_t)\Delta$ and $\avg{k_{i,t}^2}-\avg{k_{i,t}}^2=R_i(\bx_t)\Delta$. To derive the Gaussian approximation one replaces the Poissonian random variables by Gaussian noise with these first and second moments, i.e. $k_{i,t}\to R_i(\bx_t)\Delta+\sqrt{\Delta\Omega}\zeta_{i,t}$ with 
\be\label{eq:zetacorrfirst}
\avg{\zeta_{i,t}\zeta_{j,t'}}=\delta_{i,j}\delta_{t,t'}r_i(\bx_t),
\ee
and where we have used $R_i(\bx)=\Omega r_i(\bx)$. One finds
\be
x_{\alpha, t+\Delta} = x_{\alpha,t} + \Delta \sum_{i} v_{i,\alpha}r_i(\bx_t) + \sqrt{\frac{\Delta}{\Omega}}\sum_i v_{i,\alpha}\zeta_{i,t}.
\label{eq:differenceeq2}
\ee
Eq. (\ref{eq:differenceeq2}) together with Eq. (\ref{eq:zetacorrfirst}) is recognised as the Euler-Maruyama discretisation of the continuous-time CLE:
\be
\dot x_\alpha(t)=\sum_{i} v_{i,\alpha}r_i[\bx(t)]+\frac{1}{\sqrt{\Omega}}\sum_i v_{i,\alpha}\zeta_{i}(t),
\label{eq:clenodelay}
\ee
with $\avg{\zeta_{i}(t)\zeta_{j}(t')}=\delta_{i,j}\delta(t-t')r_i[\bx(t)]$, where the noise is interpreted in the It\={o} sense. This has a slightly different structure to the CLE commonly found in the literature, which is as presented in Eq.~\eqref{eq:chemlang} and Eq.~\eqref{eq:noisecorrelator}. In Eq. (\ref{eq:clenodelay}) we have one noise source, $\zeta_i$, for each reaction - similar to the conventions found in \cite{graham}. The more common choice is to have one noise source, $\eta_\alpha$, for each species. The two forms of the CLE are related via
\be
\eta_\alpha(t) = \sum_i v_{i,\alpha}\zeta_{i}(t).
\label{eq:noiseequiv}
\ee

We shall use the form of Eq.~\eqref{eq:clenodelay}, as it makes clearer the origin of the noise and is more straightforward to extend to delay reactions, both for analytical calculations and numerical simulation. In what follows we will only use the identification shown in Eq.~\eqref{eq:noiseequiv} when we make the LNA. 

\section{Gaussian approximation for chemical reaction systems with delays}
\label{sec:approxfordelay}
\subsection{Definitions and notation}
We will now introduce our notation for delay reactions. These are reactions triggered at one time, $t$, in the same way as conventional reactions without delay. Let us focus on one particular type of reaction, say $i$. As before we assume that the initiation of the reaction occurs with a rate $R_i[\bx(t)]$, and that this rate only depends on the state of the system at time $t$. Delay reactions can have an immediate effect on particle numbers, indicated by stoichiometric coefficients $v_{i,\alpha}$ as before. At the time a delay reaction of type $i$ triggers, a delay time, $\tau$, is drawn from the distribution $K_i(\cdot)$. We always imply $K_i(\tau< 0)=0$ and $\int_0^\infty d\tau K_i(\tau)=1$. At a later time, $t+\tau$, a further change of particle numbers may then occur, indicated by a second set of coefficients, $w_{i,\alpha}$.
The special case of fixed delay is recovered when $K_i(\cdot)$ is a $\delta$-distribution. Furthermore, reactions without delay are contained in this notation as well, one simply has $w_{i,\alpha} = 0$ for all $\alpha$ for such reactions, the kernel $K_i(\cdot)$ is then irrelevant. Our formalism is technically limited to reactions with delayed effects at one single subsequent time, i.e. particle numbers may change at the initial time and at most at one later time. However it is possible to extend the method to include multiple delay periods, that is reactions for which events take place at more than two distinct times.

\subsection{Derivation of the approximation}\label{sec:approx}
Similar to the Markovian case, we proceed by first discretising time into intervals of duration $\Delta$. All times $t$ and $\tau$ are then integer multiples of $\Delta$. The total number of reactions of type $i$ triggered between times $t$ and $t+\Delta$ is then a Poissonian random variable, $k_{i,t}$, with parameter $\lambda_{i,t} \equiv R_i(\bx_t)\Delta$ as before. Each of these $k_{i,t}$ reactions may have an additional effect on particle numbers at a later time. The delay periods are independently drawn from the kernel $K_i(\cdot)$ for each occurrence of the reaction. The probability that a particular delay of $\tau\in\Delta\mathbb{N}_+$ is drawn is $\Delta \times K_i(\tau)$ in the discrete-time model. The rate with which a reaction of type $i$ triggered at $t$ with a delay of precisely $\tau$ is $\lambda_{i,t}^\tau \equiv (R_i(\bx_t)\Delta)\times (\Delta K_i(\tau))$, as the triggering of the reaction and the delay period are independent. The number of reactions of type $i$ triggered at 
$t$ and with a delay of precisely $\tau$ is hence a Poissonian random variable,  $k_{i,t}^\tau$, with parameter $\lambda_{i,t}^\tau$. One has $\sum_{\tau\geq\Delta}k_{i,t}^\tau=k_{i,t}$, and the discrete-time kernels are normalised such that $\Delta\sum_{\tau\geq\Delta} K_i(\tau)=1$.

\begin{figure}[t!!]
\centerline{\includegraphics[width=0.5\textwidth]{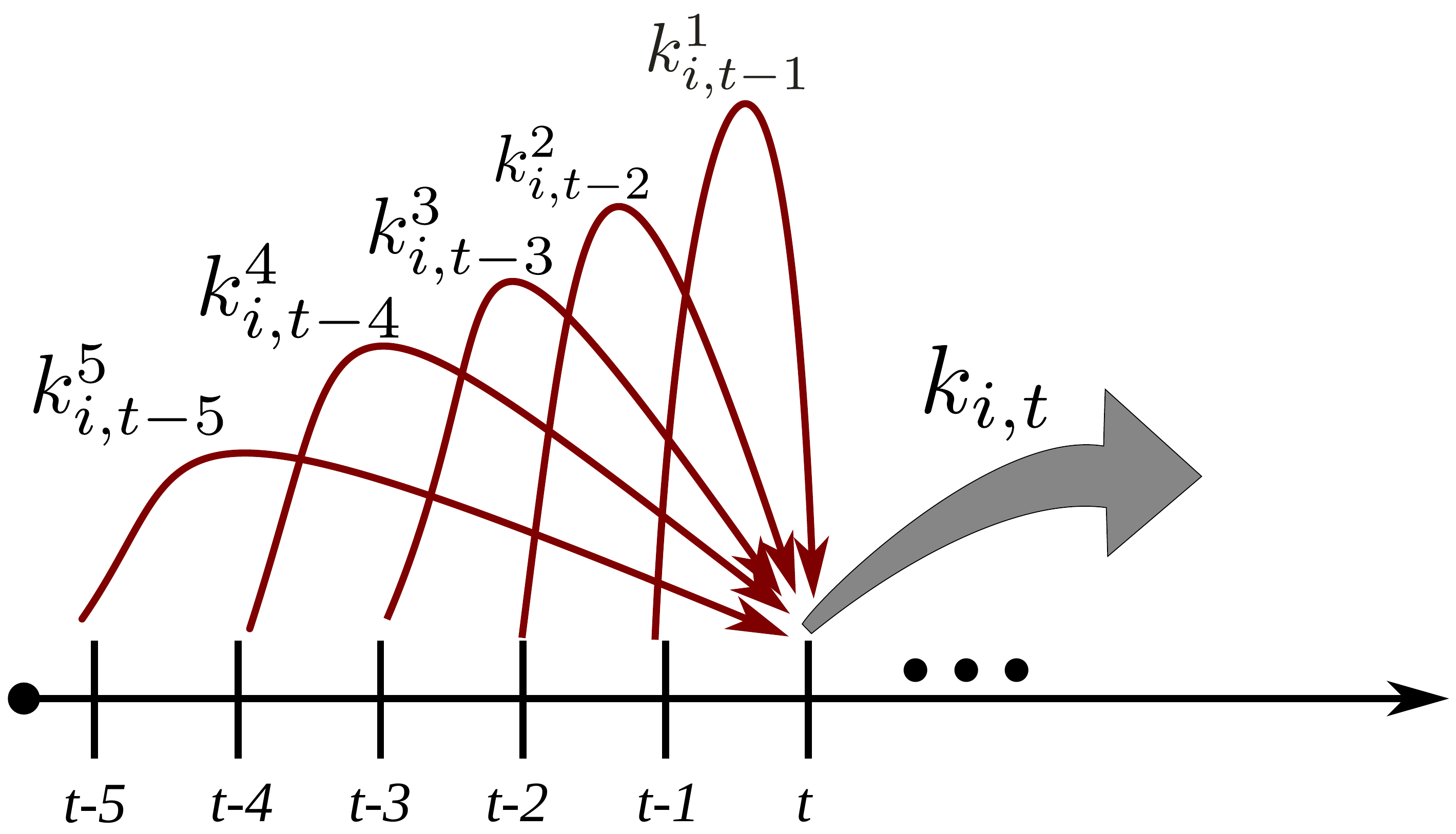}}
\vspace{0.5em}
\caption{(Colour on-line) Diagram indicating the reaction events contributing to the change of particle numbers at a given time $t$. Particle numbers are changed due to instantaneous effects of reactions triggering at time $t$, as indicated by $k_{i,t}$. These may then have delayed effects at future times, indicated by the filled arrow. Other changes of particle numbers at $t$ are due to delayed effects of reactions triggered at earlier times, as indicated by the arrows on the left. For simplicity we have set $\Delta = 1$.}
\label{fig:timeline}
\end{figure}

The change of particle concentrations at time step $t$ can then be written as
\be
x_{\alpha, t+\Delta} = x_{\alpha,t} + \frac{1}{\Omega}\left(\sum_{i} v_{i,\alpha}k_{i,t}+ \sum_i \sum_{\tau \ge \Delta} w_{i,\alpha}k_{i,t-\tau}^\tau\right).
\label{eq:delaydifference}
\ee
The first sum on the right-hand side captures the instantaneous effects of reactions at the time they are triggered. The second sum represents changes of particle numbers occurring at time $t$, but resulting from delay reactions triggered at earlier times, $t-\tau$, where $\tau=\Delta,2\Delta,\dots$. This is illustrated in Fig. \ref{fig:timeline}.

Keeping in mind that $k_{i,t}=\sum_\tau k_{i,t}^\tau$, we can write Eq. (\ref{eq:delaydifference}) as
\be
x_{\alpha, t+\Delta} = x_{\alpha,t} + \frac{1}{\Omega}\sum_{i}\sum_{\tau\geq\Delta}\left( v_{i,\alpha}k_{i,t}^\tau+ w_{i,\alpha}k_{i,t-\tau}^\tau\right).
\label{eq:delaydifference2}
\ee

Similar to the procedure in \cite{gillespie_chemlang}, the Gaussian approximation now consists in replacing the Poissonian random variables $k_{i,t}^\tau$ by Gaussian noise variables with appropriate first and second moments. The mean of the distribution of $k_{i,t}^\tau$ is $\lambda_{i,t}^\tau$, and the variance is also $\lambda_{i,t}^\tau$. We therefore replace $k_{i,t}^\tau$ as follows:
\be
k_{i,t}^\tau\to\lambda_{i,t}^\tau+ \sqrt{\Omega\Delta}\zeta_{i,t}^\tau,
\ee
where the $\zeta_{i,t}^\tau$ are independent Gaussian random variables of mean zero and with second moments $\avg{\zeta_{i,t}^\tau\zeta_{j,t'}^{\tau'}}=\delta_{i,j}\delta_{t,t'}\delta_{\tau,\tau'}\lambda_{i,t}^\tau/(\Omega\Delta)$. Recalling the definition  
$\lambda_{i,t}^\tau\equiv \Omega r_i(\bx_t)K_i(\tau)\Delta^2$ this leads to
\begin{align}
x_{\alpha, t+\Delta} =&~ x_{\alpha,t} + \Delta \sum_{i}\bigg( v_{i,\alpha}r_i(\bx_t) \nonumber \\
&+\Delta \sum_{\tau \ge \Delta} w_{i,\alpha}r_i(\bx_{t-\tau})K_i(\tau)\bigg) \nonumber \\
&+ \sqrt{\frac{\Delta}{\Omega}}\sum_{i}\sum_{\tau\geq\Delta}\left( v_{i,\alpha}\zeta_{i,t}^\tau + w_{i,\alpha}\zeta_{i,t-\tau}^\tau\right),
\label{eq:discretelangevin}
\end{align}
where 
\be\label{eq:cledcorr}
\avg{\zeta_{i,t}^\tau\zeta_{j,t'}^{\tau'}}=\delta_{i,j}\delta_{t,t'}\delta_{\tau,\tau'}r_i(\bx_t)K_i(\tau)\Delta,
\ee
and where we have used $\Delta \sum_{\tau}K_i(\tau)=1$. 
Keeping in mind that $\Delta\sum_{\tau\geq \Delta}[\dots]\rightarrow\int_0^\infty d\tau [\dots]$ in the limit of small $\Delta$, this is recognised as the discretisation of the CLE (see Appendix~\ref{appendix:contlimit} for further details)

\begin{align}\label{eq:cle}
\dot{x}_{\alpha}(t) =&~ F_\alpha[t,\bx]+ \frac{1}{\sqrt{\Omega}}\sum_i \int_0^\infty d\tau \Big\{v_{i,\alpha}\zeta_i(t,\tau) \nonumber \\
&~~~~~~~~~~~~~~~~+w_{i,\alpha}\zeta_i(t-\tau,\tau)\Big\},
\end{align}

with the so-called drift term \cite{gardiner}
\begin{align}\label{eq:drift}
F_\alpha[t,\bx] =&~\sum_{i}\bigg(\int_0^\infty d\tau~ w_{i,\alpha}r_i[\mathbf{x}(t-\tau)]K_i(\tau) \nonumber \\
&~~~~~+ v_{i,\alpha}r_i[\mathbf{x}(t)]\bigg),
\end{align}
and where 
\be\label{eq:zetacorr}
\avg{\zeta_i(t,\tau)\zeta_j(t',\tau')} = \delta_{i,j} \delta(t-t')\delta(\tau-\tau')r_i[\bx(t)]K_i(\tau).
\ee 
In this formalism reactions without delay have $w_{i,\alpha} = 0$ for all $\alpha$, as mentioned before. In this case the noise variables $\zeta_i(t,\tau)$ only enter Eq.~\eqref{eq:cle} when the reaction fires, and we can define $ \zeta_i(t) \equiv \int d\tau \zeta_i(t,\tau) $ with $\avg{\zeta_i(t)} = 0$ and $\avg{\zeta_i(t)\zeta_i(t')} = \delta(t-t')r_i[\bx(t)]$. This CLE is the starting point for further analytical approximations, and it can also be used for efficient numerical simulation of the stochastic process in the limit of weak noise, see Appendix~\ref{sec:simmethods} for details.

\subsection{Linear-noise approximation}\label{sec:lna}
The above CLE represents, in general, nonlinear dynamics with multiplicative noise. This makes analytical progress difficult, further simplifications and approximations are required. We note that the noise terms in the CLE are of order $\Omega^{-1/2}$. The dynamic variables, $x_\alpha$, are stochastic variables and they enter nonlinearly in the drift term and in the noise amplitudes. This means nonlinear effects of the intrinsic stochasticity are retained, despite the Gaussian approximation of the noise. The LNA takes the approximation one step further, and linearises the effects of noise entirely \cite{vankampen}. It retains nonlinearity only at the level of the deterministic limit; the time-evolution of fluctuations about this limit are described by a set of linear Langevin equations, containing additive noise only. The LNA is derived by separating deterministic and stochastic effects, viz. $\bx \to \bx^\infty + \Omega^{-1/2}\bxi$, and by subsequently discarding all terms of quadratic or higher order in $\Omega^{-1/2}$. The superscript $\infty$ here indicates concentrations in the deterministic limit, $\Omega\to\infty$. Applying this procedure to Eq.~\eqref{eq:cle} one finds $\dot x^\infty_\alpha = F_\alpha[t,\bx^\infty]$ on the deterministic level, i.e. to lowest order in $\Omega^{-1/2}$. At subleading order one finds that fluctuations are governed by
\be
\frac{d \xi_\alpha(t)}{d t} = \int_{-\infty}^t dt'\sum_\beta\frac{\delta F_\alpha[t,\bx^\infty]}{\delta x^\infty_\beta(t')}\xi_\beta(t') + \eta_\alpha(t),  
\ee
where $\delta F_\alpha[t,\bx^\infty]/(\delta x^\infty_\beta(t'))$ is the functional derivative of $F_\alpha[t,\bx^\infty]$ with respect to $x_\beta^\infty(t')$. We have here collected all noise terms into one single Gaussian source, $\eta_\alpha(t)$, given by
\be
\eta_\alpha(t)=\sum_i \int_0^\infty d\tau \left[v_{i,\alpha}\zeta_i(t,\tau)+w_{i,\alpha}\zeta_i(t-\tau,\tau)\right].
\ee
Within the LNA one has  $\avg{\zeta_i(t,\tau)\zeta_j(t',\tau')} = \delta_{i,j} \delta(t-t')\delta(\tau-\tau')r_i[\bx^\infty(t)]K_i(\tau)$, note the replacement $\bx(t)\to\bx^\infty(t)$ compared to the expression in Eq. (\ref{eq:zetacorr}). As explained above, in the LNA the noise becomes additive, and the correlation properties of the $\{\eta_\alpha(t)\}$ only depend on the deterministic variable $\bx^\infty(t)$. The explicit form of the correlations of the Gaussian additive noise is given by $\avg{\eta_\alpha(t)\eta_\beta(t')}=B_{\alpha,\beta}(t,t',\bx^\infty)$, where \cite{remark}
\begin{align}
&B_{\alpha,\beta}(t,t',\bx^\infty) =\sum_{i}\bigg(v_{i,\alpha}v_{i,\beta} r_i[\mathbf{x}^\infty(t)]\delta(t-t') \nonumber \\
&~~~~+\int_0^\infty d\tau~ w_{i,\alpha}w_{i,\beta} r_i[\mathbf{x}^\infty(t-\tau)] K_{i}(\tau)\delta(t-t')\nonumber \\
&~~~~+v_{i,\alpha}w_{i,\beta} r_i[\mathbf{x}^\infty(t)]K_{i}(t'-t) \nonumber \\
&~~~~+w_{i,\alpha} v_{i,\beta} r_i[\mathbf{x}^\infty(t')]K_{i}(t-t') \bigg).
\end{align}
While many analytical studies of these linearised dynamics focus on regimes in which the deterministic system reaches a fixed point (see e.g. \cite{newman_mckane}), it is important to keep in mind that the LNA can be derived for more general classes of model systems. In the general case the resulting Langevin equation has time-dependent coefficients, derived from the underlying deterministic trajectory. Progress can be made not only when expanding about fixed points, but also when the deterministic system approaches a limit cycle \cite{boland, constable}. Recently, expansion techniques have also been applied to chaotic systems in discrete time \cite{challenger}, although there is then relatively little scope for a full analytical characterisation. The LNA does have some limitations, and can not be used for all problems. For example it breaks down when the system is not monostable, when stochasticity induces transitions between attractors, and when non-linear effects are important (an example of which will be studied in Sec.~\ref{sec:bruss_spike}).

\section{Stochastic Brusselator with delay}\label{sec:bruss}
\subsection{Model definition}

The Brusselator is a paradigmatic model for oscillations in chemical reaction systems \cite{prigogine,lefever, brown.1995, gray}. Previous work on stochastic versions of the Brusselator model include the study of fluctuations about limit cycles \cite{boland}, pattern formation in spatial variants \cite{biancalani}, and a recent study on noise-induced switching between large and small amplitude oscillations \cite{giver}. 

We here study a non-Markovian variation of the Brusselator model. This variant is not chosen with a particular real-world chemical system in mind, but instead in order to demonstrate how the techniques discussed in the previous section apply to a specific example. Our choice of model is motivated by the work of \cite{woolley}, who have studied delay reactions of the same form in the context of pattern formation in developmental biology. This existing work, however, is based on numerical simulations only, no analytical calculations are presented in \cite{woolley}. 

The reactions by which we define the Brusselator with delay are the following:

\BE
2X_1+X_2 &\overset{c}{\longrightarrow}& 2X_1+ Y;  Y \overset{K(\tau)}{\longrightarrow} X_1, \nonumber\\
\emptyset &\overset{a}{\longrightarrow}& X_1, \nonumber\\
X_1 &\overset{1}{\longrightarrow}& \emptyset, \nonumber\\
X _1&\overset{b}{\longrightarrow}& X_2.
\label{eq:Bruss_reactions}
\EE
The reactions are indexed with $i = 1,2,3,4$ from top to bottom. We here focus on a well-mixed reactor. The first reaction is a delay reaction, and is the sole distinction between this model and the conventional Brusselator. The above notation indicates a two-step reaction: at the time the reaction is triggered two molecules of type $X_1$ react with one particle of type $X_2$, to generate a particle of type $Y$. The $X_2$-particle is removed in this process. Following mass-action principles, reactions of this type are triggered with a rate $cn_1^2n_2/\Omega^2$,where $c$ is a constant, and where $n_1$ and $n_2$ indicate the number of particles of types $X_1$ and $X_2$ in the reaction container respectively. As before $\Omega$ sets the scale of the system size. The intermediate particle $Y$ remains for a duration, $\tau$, drawn from a distribution $K(\cdot)$. The second segment of the delay reaction occurs $\tau$ units of time later, 
when the $Y$-particle decays into a particle of type $X_1$. The remaining reactions are standard, and represent the creation and removal of particles of type $X_1$ (second and third reaction above, respectively), and the conversion of particles of type $X_1$ into particles of type $X_2$ (fourth reaction).

\subsection{Chemical Langevin equation and deterministic limit}
The CLE obtained from reactions Eq.~\eqref{eq:Bruss_reactions} and using the above formalism are
\begin{widetext}
\BE
\frac{d x_1(t)}{dt} &=& a - (1+b)x_1(t) + c\int_{-\infty}^td\tau ~K(t-\tau)x_1(\tau)^2x_2(\tau) \nonumber \\
&&+ \frac{1}{\sqrt{\Omega}}\left\{\zeta_2(t) -\zeta_3(t) - \zeta_4(t)+ \int_0^\infty d\tau~ \zeta_1(t-\tau,\tau)\right\}, \nonumber\\
\frac{d x_2(t)}{dt} &=& b x_1(t) - cx_1(t)^2x_2(t) + \frac{1}{\sqrt{\Omega}} \left\{\zeta_4(t)-\int_0^\infty d\tau~\zeta_1(t,\tau)\right\},\nonumber \\
\frac{dy(t)}{dt}&=&cx_1(t)^2x_2(t)- c\int_{-\infty}^td\tau ~K(t-\tau)x_1(\tau)^2x_2(\tau)+ \frac{1}{\sqrt{\Omega}}\int_0^\infty d\tau \left\{\zeta_1(t,\tau)-\zeta_1(t-\tau,\tau)\right\} .
\label{eq:bruss_chemlang}
\EE
\end{widetext}
The noise correlators are 
\begin{align}
\avg{\zeta_1(t,\tau)\zeta_1(t',\tau')} =&~ \delta(t-t')\delta(\tau-\tau') \nonumber \\
&~~\times cx_1(t)^2x_2(t)K(\tau), \nonumber \\
\avg{\zeta_2(t)\zeta_2(t')} =&~ \delta(t-t')a, \nonumber \\
\avg{\zeta_3(t)\zeta_3(t')} =&~ \delta(t-t')x_1(t), \nonumber \\
\avg{\zeta_4(t)\zeta_4(t')} =&~ \delta(t-t')bx_1(t).\label{eq:bruss_correl}
\end{align}

Note that the subscripts of $\zeta_i$ refer to the reaction number and the subscripts of $x_\alpha$ refer to the species. We notice that the dynamics for $x_1$ and $x_2$ is independent of that for $y$, in the sense that the first two equations in Eq.~(\ref{eq:bruss_chemlang}) along with Eq.~\eqref{eq:bruss_correl} constitute a closed set of equations. This is because the concentration of particles of type $Y$ does not enter into any of the reaction rates. Given this decoupling, we will focus on the dynamics of $x_1$ and $x_2$ from this point forward.

In the deterministic limit, $\Omega\to\infty$, the noise terms in Eq.~\eqref{eq:bruss_chemlang} vanish, and one obtains the following deterministic equations,
\begin{align}
\frac{d x_1^\infty(t)}{dt} =&~  c\int_{-\infty}^td\tau ~K(t-\tau)x_1^\infty(\tau)^{2}x_2^\infty(\tau) \nonumber \\
&+a - (1+b)x_1^\infty(t), \nonumber \\
\frac{d x_2^\infty(t)}{dt} =&~ b x_1^\infty(t) - cx_1^\infty(t)^2x_2^\infty(t).
\label{eq:bruss_meanfield}
\end{align}
As before, the superscript $\infty$ is used to indicate that these equations apply to concentrations evaluated in the limit of an infinite system. This deterministic system has a unique fixed point at
\BE
x_1^* &=& a, \nonumber \\
x_2^* &=& b/(ac),
\label{eq:brussfp}
\EE
identical to that of the conventional Brusselator system \cite{lefever}. For systems without delay it is straightforward to characterise the stability of fixed points from the eigenvalues of the corresponding Jacobian. Often the corresponding phase diagram can then be computed analytically. For systems with delay such an analysis is more involved. Delay kernels lead to transcendental equations for the equivalent of eigenvalues, and these typically have an infinite number of solutions in the complex plane. The assessment of the local stability for a fixed point of a delay system is hence mostly limited to numerical methods \cite{ Michiels}.

\subsection{Linear-noise approximation}
We will now proceed to make further analytical progress towards describing the stationary state of the stochastic delay Brusselator system. Our starting point is the general expression for the LNA for delay systems in Sec. \ref{sec:lna}. For the Brusselator system with delay one finds
\begin{align}
\frac{d \xi_1(t)}{dt} =&~- (1+b)\xi_1(t) \nonumber \\
&+ c\int_{-\infty}^td\tau ~K(t-\tau)\bigg\{2x_1^\infty(\tau)x_2^\infty(\tau)\xi_1(\tau) \nonumber \\
&~~~~~~~~~~~~~~~~+ {x_1^\infty}(\tau)^2\xi_2(\tau)\bigg\} + \eta_1(t),\nonumber\\
\frac{d \xi_2(t)}{dt} =&~  b\xi_1(t) - 2cx_1^\infty(t)x_2^\infty(t)\xi_1(t) \nonumber \\
&- c{x_1^\infty(t)}^2\xi_2(t) + \eta_2(t),
\label{eq:lna_langevin}
\end{align}
with noise correlators $\avg{\eta_\alpha(t)\eta_\beta(t')}=B_{\alpha\beta}(t,t',\bx^\infty)$ given by 
\begin{align}
&B_{11}(t,t',\bx^\infty) =~ \delta(t-t')\bigg\{ a + (1+b)x^\infty_1(t) \nonumber \\
&~~~~~~~~~~~~~+ c\int_{-\infty}^t d\tau ~K(t-\tau)[x^\infty_1(\tau)]^2x^\infty_2(\tau) \bigg\} , \nonumber \\
&B_{22}(t,t',\bx^\infty) =\delta(t-t')\left\{b x^\infty_1(t) + c[x^\infty_1(t)]^2x^\infty_2(t)\right\}, \nonumber \\
&B_{12}(t,t',\bx^\infty) = -cK(t-t')[x^\infty_1(t')]^2x^\infty_2(t').
\label{eq:brussnoise}
\end{align}
 We focus on a parameter regime in which the deterministic system approaches the fixed point $(x_1^*,x_2^*)$ given in Eq.~\eqref{eq:brussfp}. At asymptotic times the LNA then consists of a set of linear Langevin equations with constant coefficients, which makes further analysis particularly straightforward.  One has
\begin{align}
\frac{d \xi_1(t)}{dt} =&\int_{-\infty}^td\tau ~K(t-\tau)\bigg\{2b\xi_1(\tau) + a^2c\xi_2(\tau)\bigg\} \nonumber \\ 
&- (1+b)\xi_1(t) + \eta_1(t),\nonumber\\
\frac{d \xi_2(t)}{dt} =&  -b\xi_1(t) - a^2c\xi_2(t) + \eta_2(t), \label{eq:bruss_lna2}
\end{align}
with  
\be
 B(t,t',\bx^*) = \left(
\begin{array}{cc}
2a(1+b)\delta(t-t')&-abK(t-t')\\ -abK(t'-t) & 2ab\delta(t-t') 
\end{array}
\right).
\label{eq:lna_correlators}
\ee
From these expressions, a complete statistical characterisation of the stationary state of the linearised dynamics can be obtained. The power spectra, $S_{\alpha\beta}(\omega) = \avg{\widetilde\xi_\alpha(\omega)\widetilde\xi_\beta (-\omega)}$, of fluctuations about the deterministic fixed point can be found after Fourier transforming Eq.~\eqref{eq:bruss_lna2} and Eq.~\eqref{eq:lna_correlators} \cite{gardiner,footnote1}. The elements of the matrix $S(\omega)$ are the Fourier transforms of the cross-correlation and auto-correlation functions of the $\{\xi_\alpha(t)\}$, $C_{\alpha\beta}(\tau)=\avg{\xi_\alpha(t)\xi_\beta(t+\tau)}$. The equal-time covariance matrix $\Xi_{\alpha\beta}= \avg{\xi_\alpha(t)\xi_\beta(t)}$ in the stationary state is found as $\Xi_{\alpha\beta} = C_{
\alpha\beta}(0)=  \int_{-\infty}^\infty d\omega S_{\alpha\beta}(\omega)/(2\pi)$. The steady-state probability distribution $P^s(\bx)$ for the concentration vector $\bx$ is then given by the multi-variate Gaussian
\be
P^s(\bx) = \frac{\Omega^{1/2}}{2\pi |\mbox{det}~\Xi|^{1/2}}e^{-\Omega(\bx-\bx^*)^T\Xi^{-1}(\bx-\bx^*)/2}.
\label{eq:ss_probdist}
\ee
In these calculations the only difference between the delay case and the well-studied non-delay case is the fact that the expression for $S(\omega)$ is more complicated, in particular it involves the Fourier transform of the delay kernel $K(\cdot)$. For the Brusselator model with delay we find $A(\omega) \widetilde\bxi(\omega) = \widetilde \boldeta(\omega)$, where the matrix $A(\omega)$ reads
\be
A(\omega) = \left(\begin{array}{cc}
i\omega + (1+b) -  2b\widetilde K(\omega)&  -a^2c\widetilde K(\omega) \\ b& i\omega + a^2c 
\end{array}
\right).
\ee 
From this one has
\be\label{eq:spectra}
S(\omega) = A^{-1}(\omega)B(\omega)(A^{\dagger})^{-1}(\omega),
\ee
 where $\dagger$ denotes the conjugate transposition and where
\be
 B(\omega) = \left(
\begin{array}{cc}
2a(1+b)&-ab\widetilde K(\omega)\\ -ab\widetilde K^*(\omega) & 2ab 
\end{array}
\right).
\label{eq:ft_lna_correlators}
\ee
It follows, for example, that
\be
S_{11}(\omega) = \frac{2a(1+b)(a^4c^2+\omega^2)}{|\det A(\omega)|^2}.
\label{eq:brussps}
\ee
Similar expressions can be found for $S_{22}(\omega)$ or indeed for cross-correlation spectra. Eq.~\eqref{eq:brussps} can be numerically integrated (over $\omega$) to find $\Xi_{11}$ and hence $P^s(x_1)$. 

\subsection{Comparison with numerical results}

\subsubsection{Quasi-cycles and stationary distributions}
\begin{figure}[t!!]
\centerline{\includegraphics[angle =270,width=0.5\textwidth]{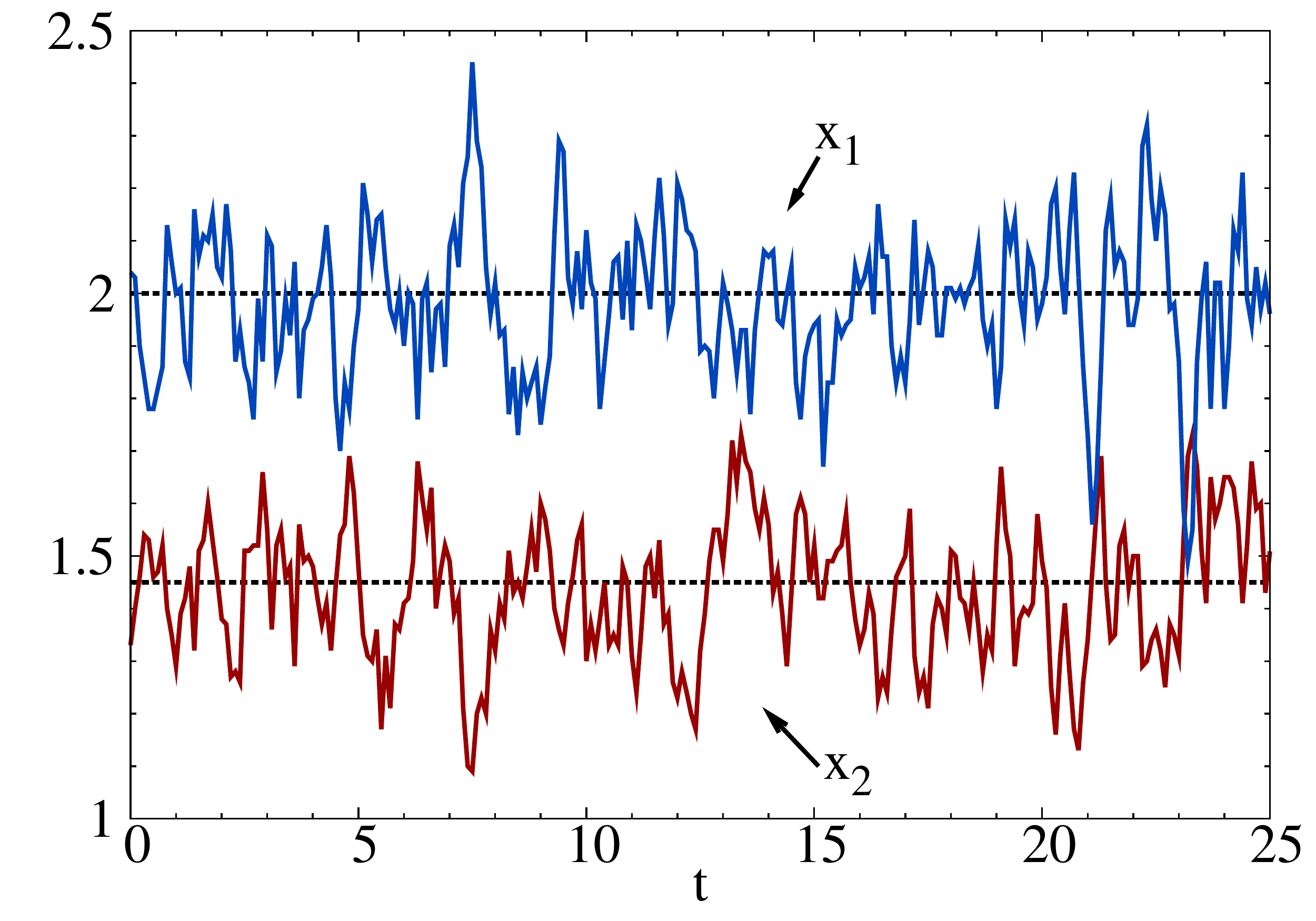}}
\vspace{0.5em}
\caption{(Colour on-line) Sample trajectory of the delay Brusselator with fixed delay. Parameters are $\Omega = 100$,~ $a = 2$,~$b = 2.9$,~$c = 1$, $\tau = 2$. Oscillations with period of approximately $2$ time units can be identified visually. These observations are validated by the results shown in Fig.~\ref{fig:bruss_ps_kernels}. The system has been initialised at the deterministic fixed point $x_1 =2$, $x_2=1.45$.}
\label{fig:bruss_trajectory}
\end{figure}

\begin{figure}[t!!]
\centerline{\includegraphics[angle =270,width=0.5\textwidth]{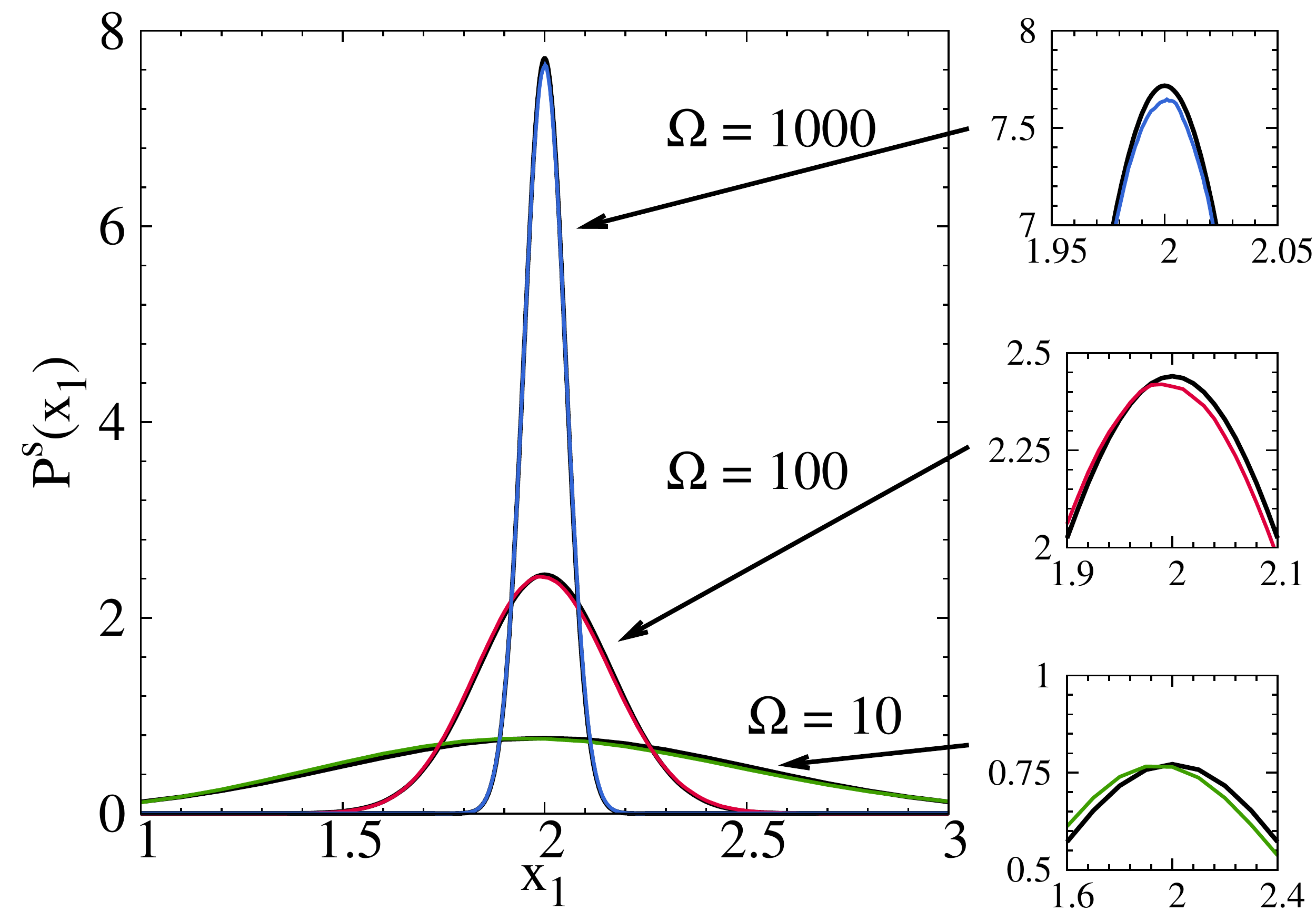}}
\vspace{0.5em}
\caption{(Colour on-line) Marginal steady-state probability distribution $P^s(x_1)$ for the Brusselator with fixed delay and for different system sizes. Parameters are $a = 2$, $b = 2.9$, $c = 1$, and $\tau = 2$. Results obtained from the LNA (black lines) provide a good approximation to the exact simulations of the process using the MNRM, shown using coloured (grey) lines. }
\label{fig:bruss_probdistdelta}
\end{figure}

A sample trajectory for the Brusselator with delay kernel $K(t) = \delta(t - \tau)$ is shown in Fig.~\ref{fig:bruss_trajectory}. The data is from numerical simulations using the modified next-reaction method (MNRM) \cite{anderson}. The MNRM is an algorithm which simulates the stochastic process exactly, for details see Appendix~\ref{sec:simmethods}. Model parameters are chosen such that the limiting deterministic model reaches a stable fixed point. The trajectory is shown in the steady state and, as seen in the figure, it fluctuates about the deterministic fixed point. The corresponding marginal distribution $P^s(x_1)$ in the steady state is shown in Fig.~\ref{fig:bruss_probdistdelta} for different choices of the system size $\Omega$. Semi-analytical results for this distribution, calculated using Eq.~\eqref{eq:ss_probdist}, are in good agreement with the results of the numerical simulations, even for the relatively low value of $\Omega=10$. We stress that the LNA is only valid when effects of higher order than $\Omega^{-1}$ can be safely neglected \cite{gardiner}. As expected, the distribution $P^s(x_1)$ becomes more sharply peaked around the deterministic fixed point when the system size is increased.
\\

Another feature seen  in Fig.~\ref{fig:bruss_trajectory} is a temporal structure to the fluctuations. The concentration variables appear to undergo noisy oscillations with an average period of approximately $T=2$. Recall that simulations are carried out with parameter values for which the deterministic system has a stable fixed point. Hence the oscillations are an effect of the intrinsic stochasticity, they are noise-driven quasi-cycles which have been widely discussed for other systems in the literature, see e.g. \cite{newman_mckane}. These cycles can be analysed further by means of the power spectrum $S_{11}(\omega)$, similar to existing studies of delay systems \cite{galla, brett_galla}.

\begin{figure}[t!!]
\centerline{\includegraphics[width=0.5\textwidth]{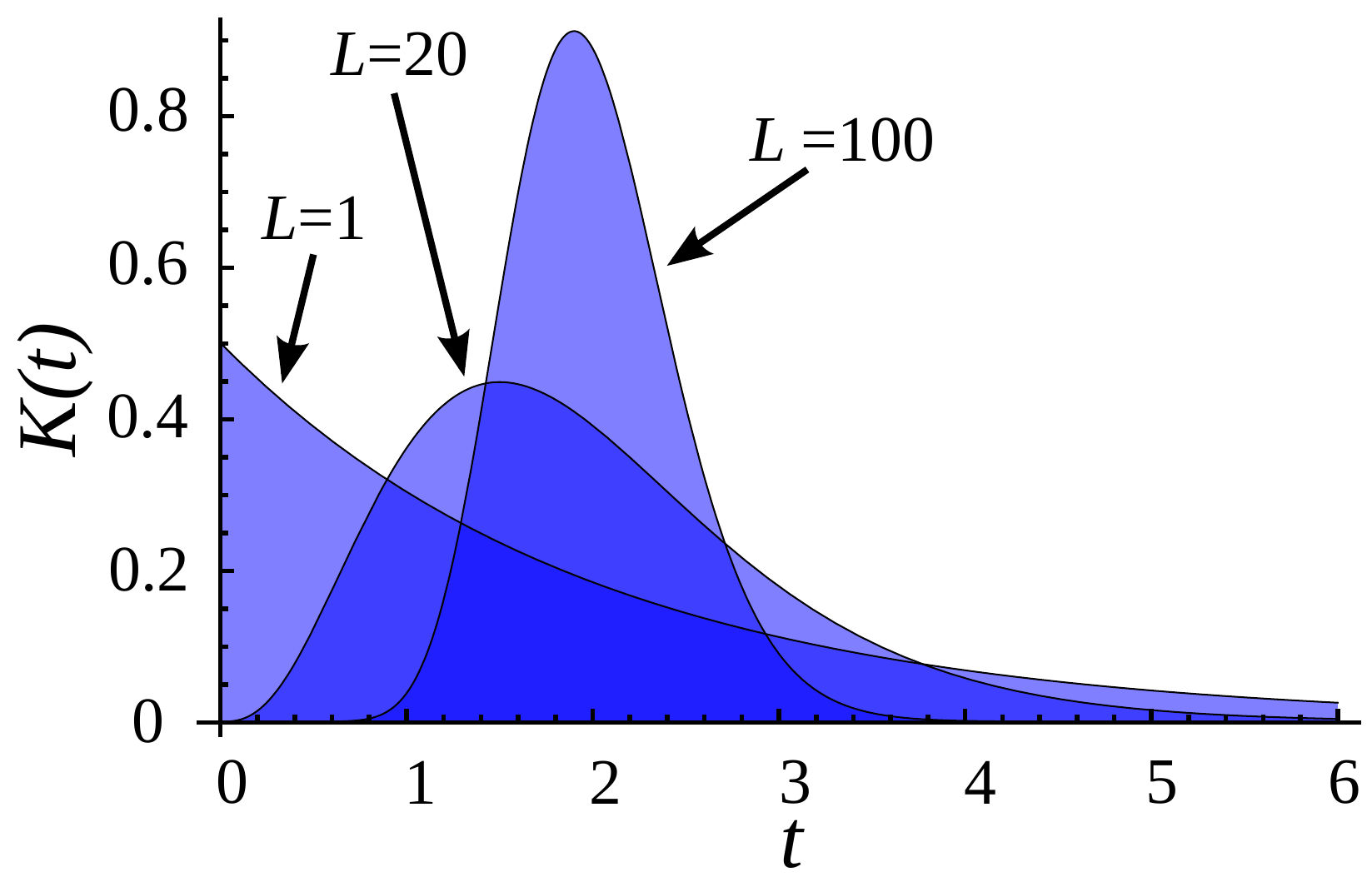}}
\vspace{0.5em}
\caption{(Colour on-line) Illustration of the $\Gamma$-distributed delay kernels, Eq. (\ref{eq:gamma}), used for the simulations in Fig. \ref{fig:bruss_ps_kernels}. The average delay is fixed to $\tau=2$.}
\label{fig:bruss_delaydist}
\end{figure}

While Eq. (\ref{eq:spectra}) provides a general expression, applicable to arbitrary delay kernels, we carry out this analysis for a family of $\Gamma$-distributed delays with a fixed average delay $\tau$. Specifically we choose
\be\label{eq:gamma}
K(t) = \frac{(L/\tau)^L}{\Gamma(L)}t^{L-1}e^{-Lt/\tau},
\ee
where $\Gamma(L)=\int_0^\infty du ~ u^{L-1}e^{-u}$. The parameter $L$ controls the shape of the distribution, as shown in Fig.~\ref{fig:bruss_delaydist}. The kernel becomes increasingly sharply distributed about $\tau$ as $L$ is increased. For $L = 1$ the kernel $K(t)$ is an exponential distribution, and for $L \to \infty$ it is the $\delta$-distribution $K(t) = \delta(t-\tau)$ used in Figs.  \ref{fig:bruss_trajectory} and \ref{fig:bruss_probdistdelta}. For all $L$ one has $\int_0^\infty dt ~tK(t)=\tau$.

\begin{figure}[t!!]
\centerline{\includegraphics[angle =270,width=0.5\textwidth]{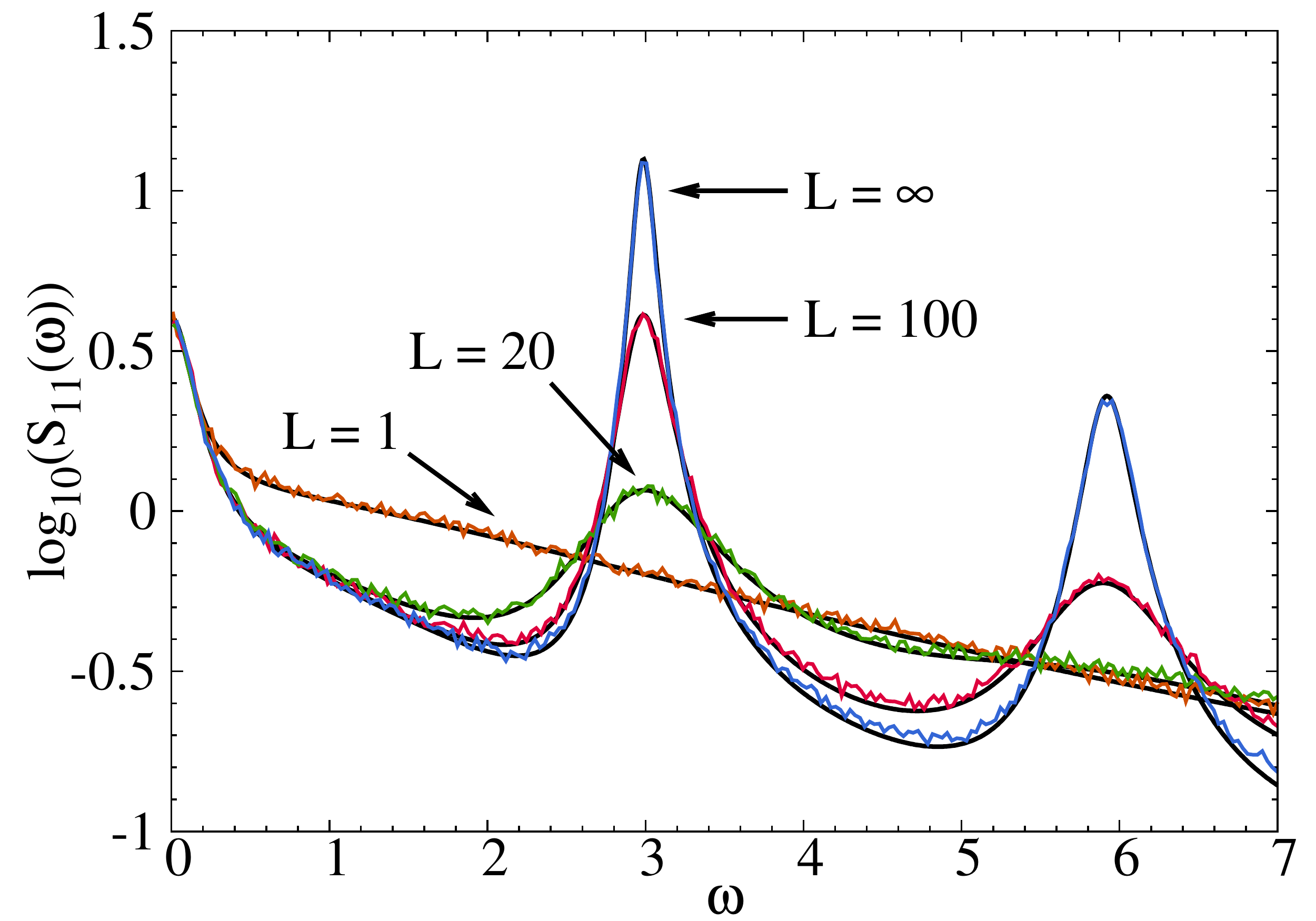}}
\vspace{0.5em}
\caption{(Colour on-line) Power spectra $S_{11}(\omega)$ of the delay Brusselator for $\Gamma$-distributed delay kernels (see Eq. (\ref{eq:gamma})) with $\tau=2$ and $L= 1, 20, 100, \infty$. Noise-induced oscillations are observed for large values of $L$, but not for the exponential delay kernel, $L=1$. The remaining parameters are the same as in Fig.~\ref{fig:bruss_probdistdelta}. Noisy lines show data from simulations using the MNRM with $\Omega = 100$, smooth lines are from the theory, Eq. (\ref{eq:brussps}).}
\label{fig:bruss_ps_kernels}
\end{figure}

Results for the power spectrum, $S_{11}(\omega)$, are shown in  Fig.~\ref{fig:bruss_ps_kernels}  for different choices of $L$. In all cases we find very good agreement between the theoretical predictions and simulation data. For $L=1$ the power spectrum of fluctuations in the concentration of particles of type $X_1$ decreases monotonically from its peak value at $\omega = 0$. This is characteristic of a system which does not display noise-induced oscillations. For larger values of $L$, i.e. more sharply peaked delay distributions, an additional peak emerges at $\omega\approx 3$ corresponding to oscillations with period of approximately $T=2.1$. Higher harmonics are seen as well. As $L \to \infty$ this peak grows and becomes dominant, indicating coherent stochastic oscillations and confirming the observations of Fig.~\ref{fig:bruss_trajectory}. The steady-state probability distribution $P^s(x_1)$ as a function of $L$ can also be investigated. It turns out that the variation with $L$ is relatively insignificant, so that we do not show results here, and limit ourselves to stating that the semi-analytical theory predicts the simulation outcome to a good accuracy.

\begin{figure*}[t!!]
\centerline{\includegraphics[angle =270,width=0.3\textwidth]{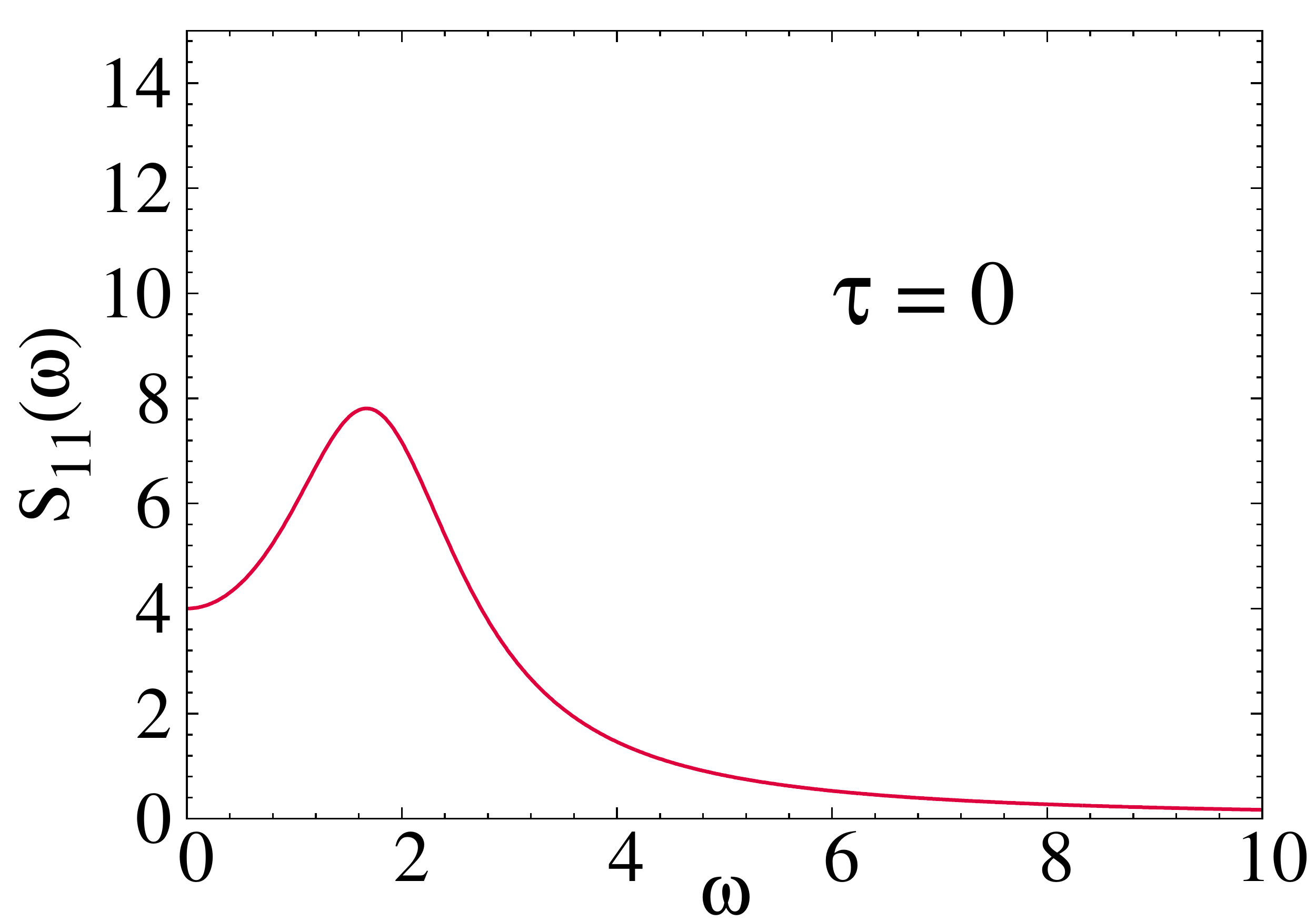}
\includegraphics[angle =270,width=0.3\textwidth]{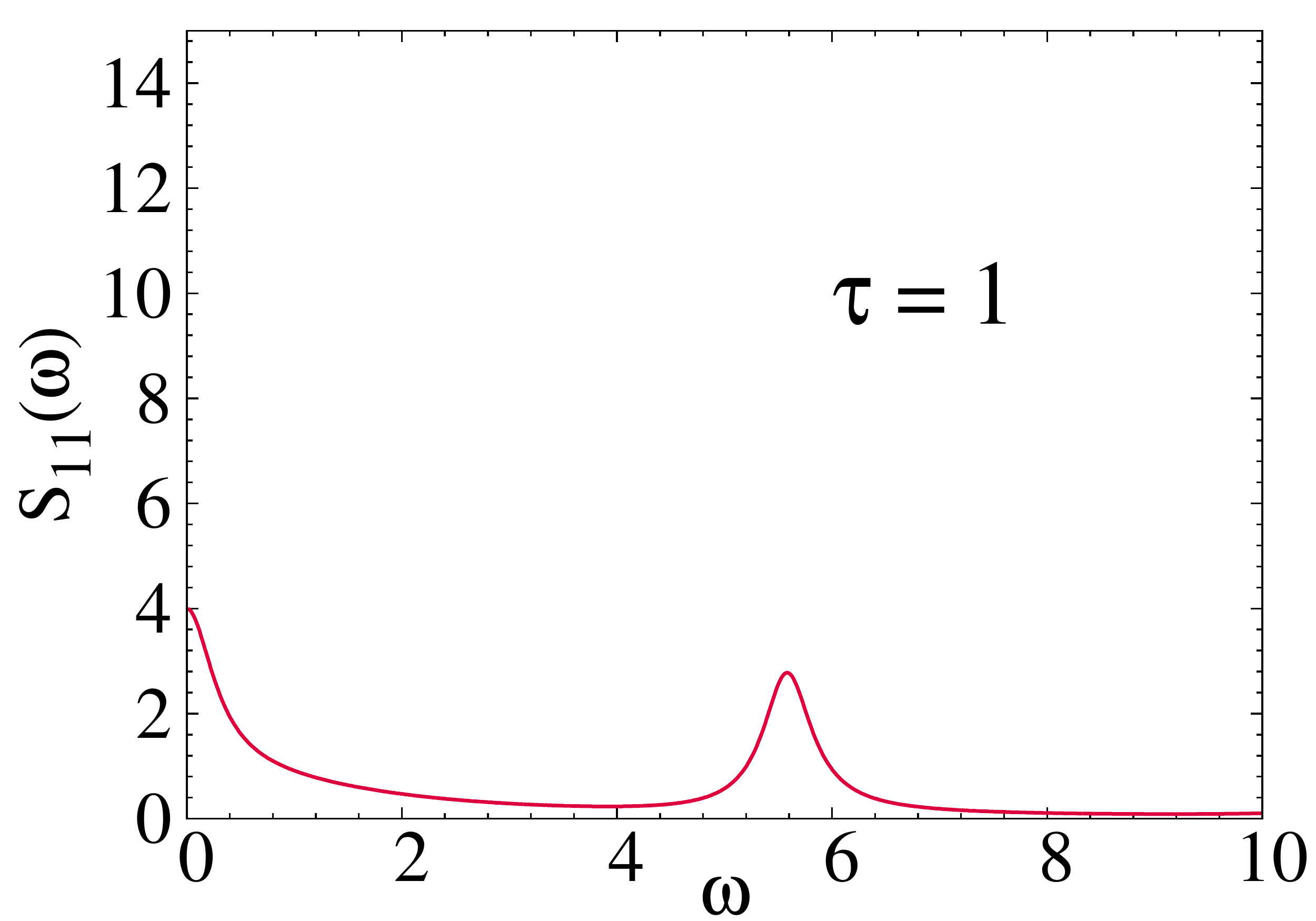}
\includegraphics[angle =270,width=0.3\textwidth]{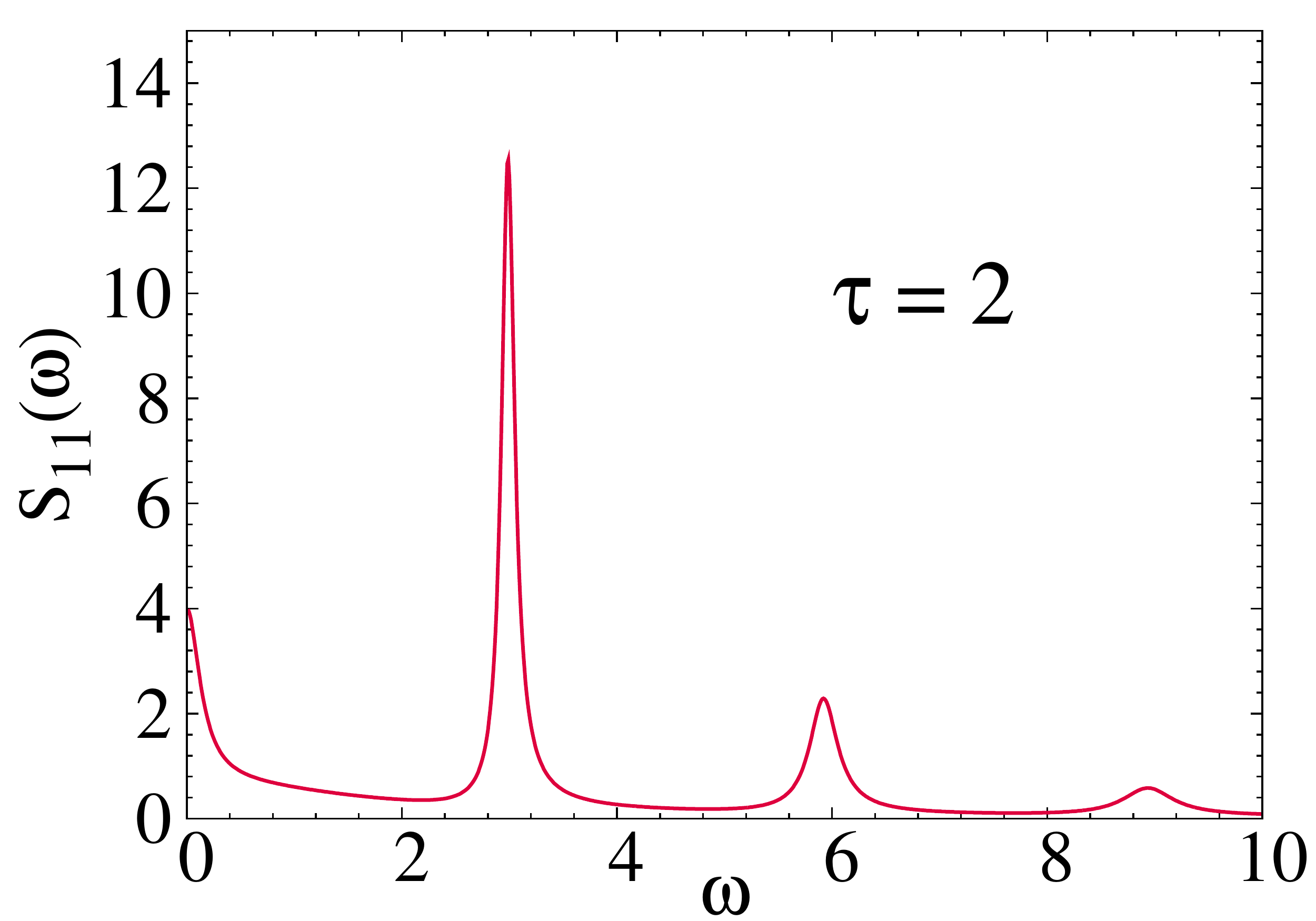}}
\centerline{
\includegraphics[angle =270,width=0.3\textwidth]{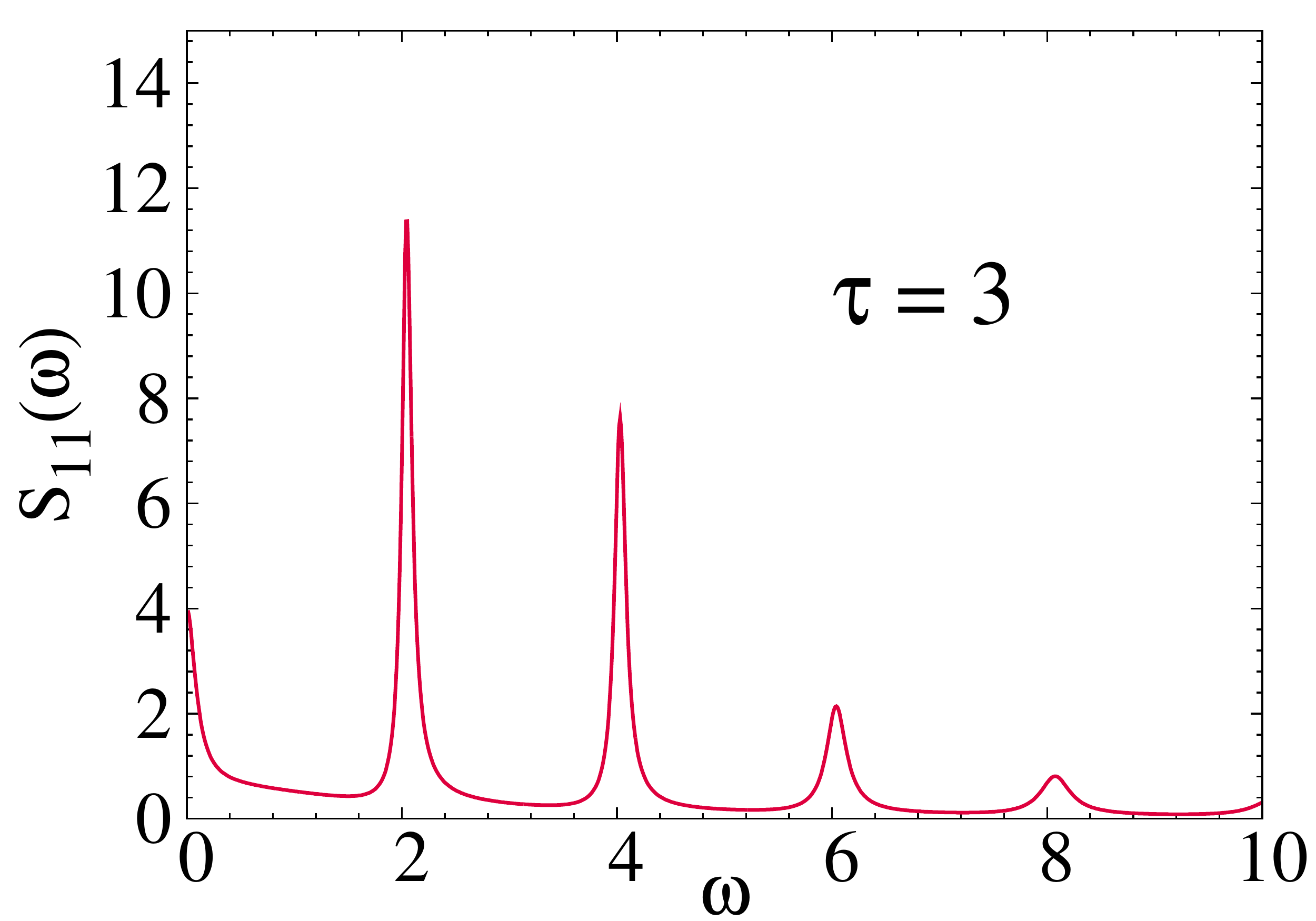}
\includegraphics[angle =270,width=0.3\textwidth]{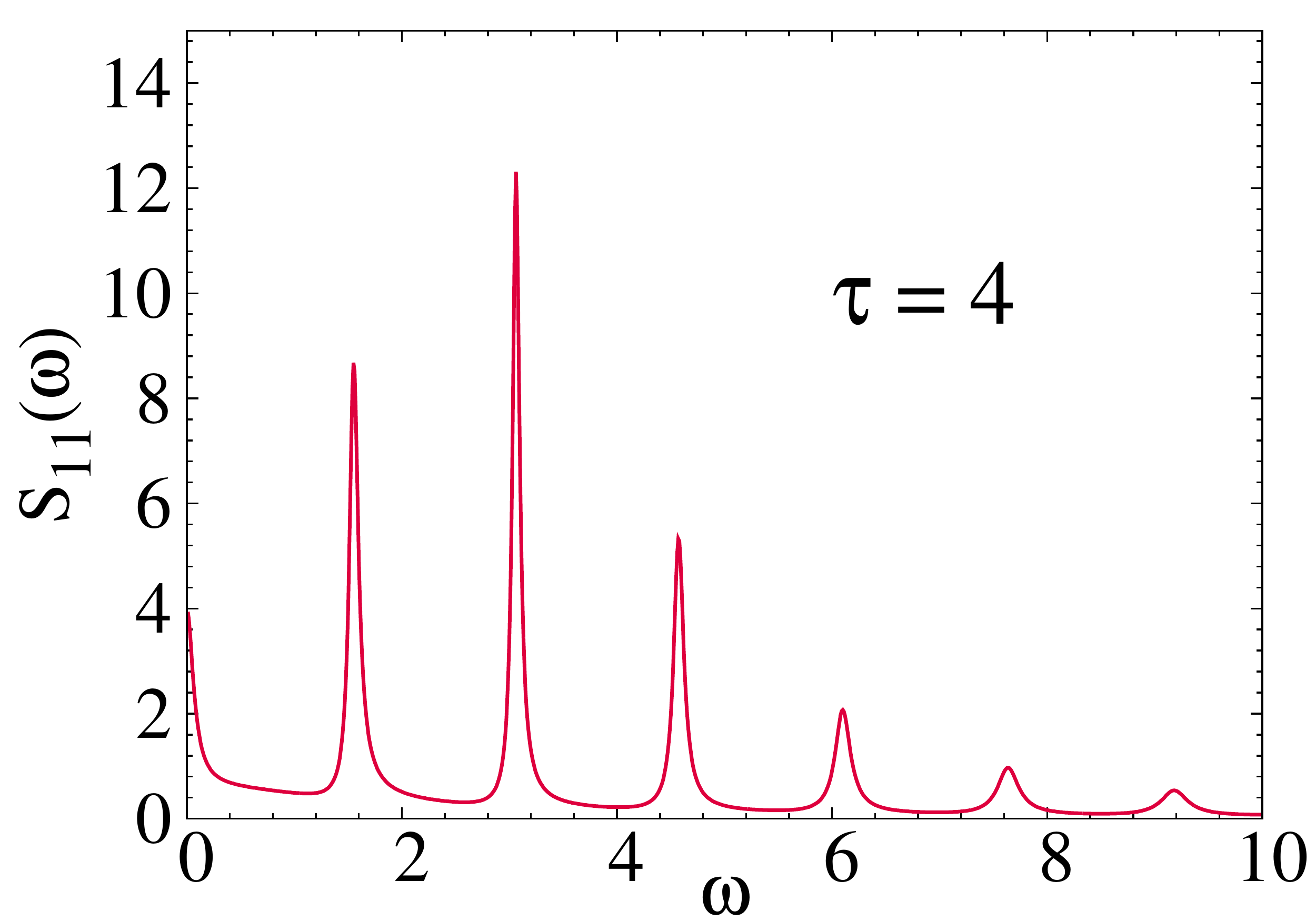}
\includegraphics[angle =270,width=0.3\textwidth]{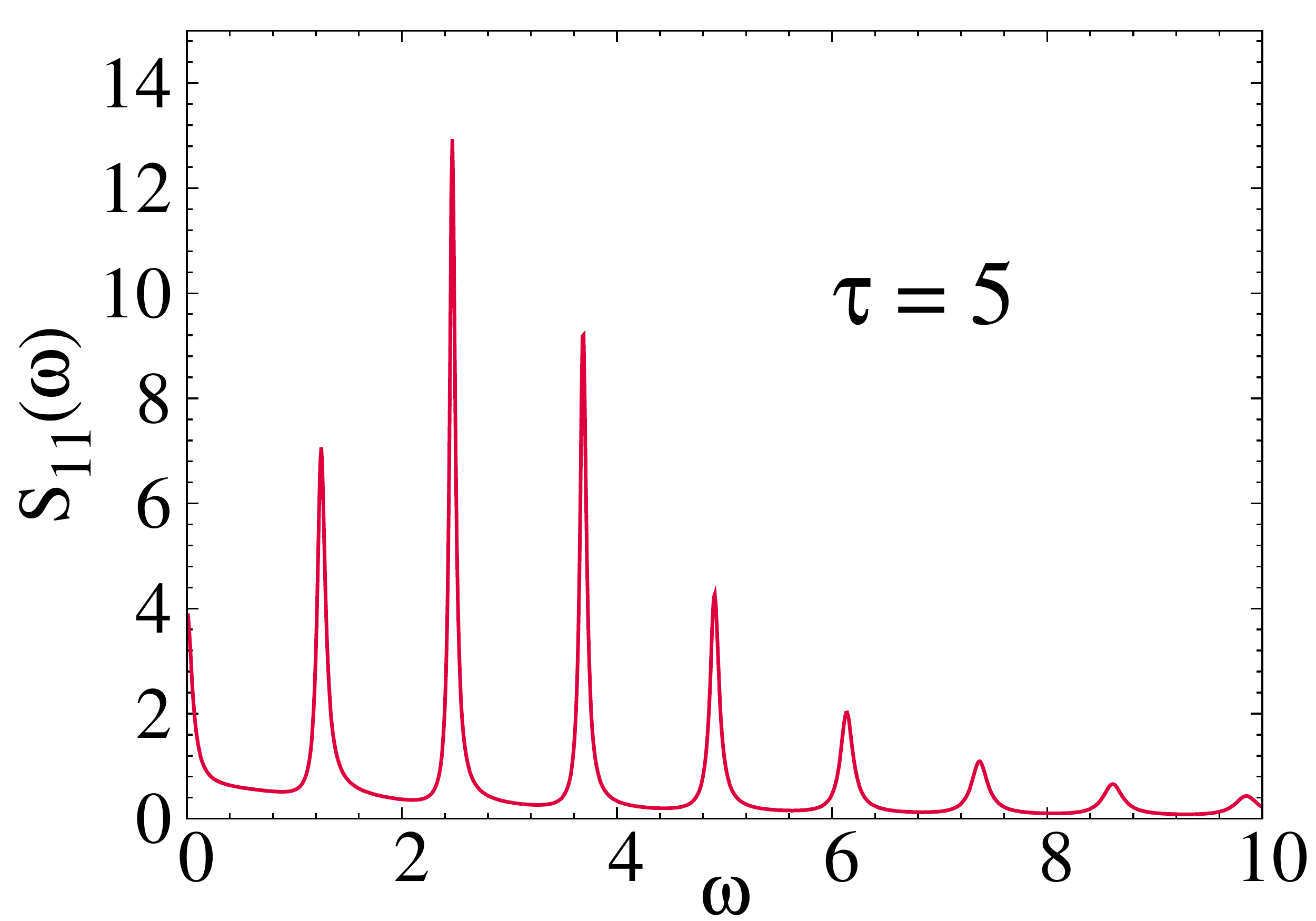}
}
\vspace{0.5em}
\caption{(Colour on-line) Power spectra $S_{11}(\omega)$ for the delay Brusselator, calculated from Eq.~\eqref{eq:brussps} with fixed delay $\tau$. The remaining parameters are the same as in Fig.~\ref{fig:bruss_probdistdelta}. Exact simulations with the MNRM have also been performed to confirm the spectra (data not shown).}
\label{fig:spectrafixeddelay}
\end{figure*}

\begin{figure}[t!!]
\centerline{\includegraphics[angle =270,width=0.5\textwidth]{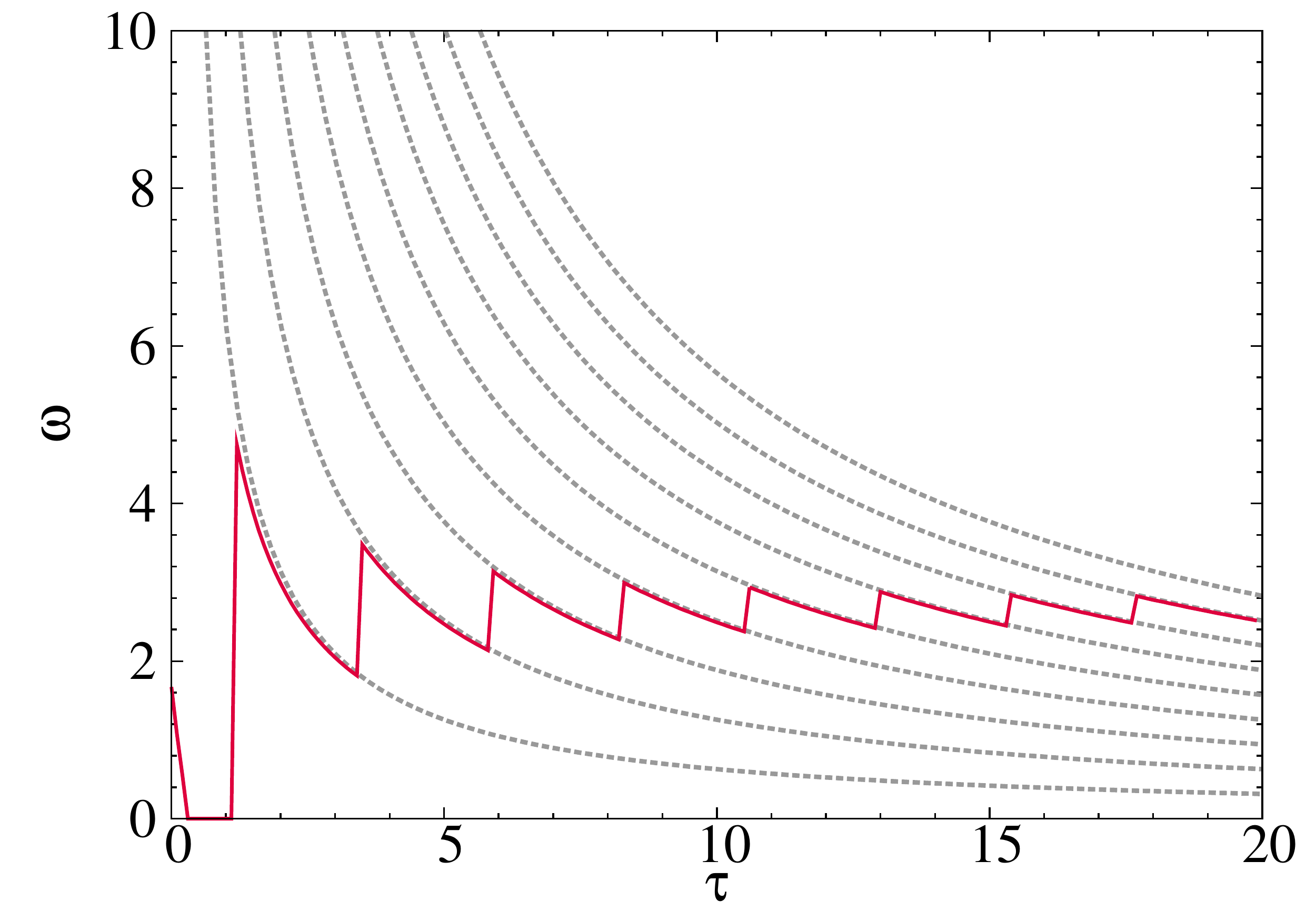}}
\vspace{0.5em}
\caption{(Colour on-line) Angular frequency $\omega$ at which the power spectrum $S_{11}(\omega)$ attains its global maximum in the Brusselator with fixed delay $\tau$. The red (solid) line is calculated from Eq.~\eqref{eq:brussps}. The grey (dashed) lines correspond to $\omega_n(\tau)= \frac{2\pi n}{\tau}$ ($n=1,\dots,9$ are shown). Model parameters other than $\tau$ are the same as in Fig.~\ref{fig:bruss_probdistdelta}.}
\label{fig:omegamax}
\end{figure}

\subsubsection{Effects of the delay period}

The effects of varying the delay period, $\tau$, on the power spectrum of the delay Brusselator with fixed delay are shown in Figs.~\ref{fig:spectrafixeddelay} and \ref{fig:omegamax}. When there is no delay (top-left panel of Fig.~\ref{fig:spectrafixeddelay}) the power spectrum of fluctuations in the conventional Brusselator is recovered \cite{boland}. For the model parameters $a,b$ and $c$ chosen here, the system displays noise-induced cycles, as indicated by the peak at a non-zero frequency. As $\tau$ is increased this peak shrinks and disappears. New peaks appear located approximately at frequencies $\omega_n=2n\pi/\tau$, with $n=1,2,\dots$. As $\tau$ is increased the peak located at $\omega_n$ grows and moves to the left, c.f. the first peak in the top-middle and top-right panels of Fig.~\ref{fig:spectrafixeddelay}. We find empirically that the peak height of the $n$-th peak increases with $\tau$ so long as $\omega_n\gtrsim 2.6$. Once $\tau$ is sufficiently large so that $\omega_n \lesssim 2.6$ the height of peak $n$ decreases upon further increase of $\tau$, see e.g. height of the first peak in the spectra for $\tau=2, 3, 4, 5$ in Fig. \ref{fig:spectrafixeddelay}.

Fig.~\ref{fig:omegamax} shows the value of $\omega$ for which $S_{11}(\omega)$ is at its maximum. For small $\tau$ this maximum is attained at a small angular frequency, and then jumps to a significantly larger value of $\omega$ at $\tau\approx 1.2$. This occurs when the maximum of $S_{11}(\omega)$ is attained at $\omega\approx \omega_1=2\pi/\tau$. Further increasing $\tau$ the maximum of the power spectrum remains at the first peak. The location of this peak, $\omega\approx 2\pi/\tau$, gradually decreases along with peak height. At $\tau\approx 3.5$ the second peak becomes dominant, and the maximum of the spectrum is now found at $\omega\approx\omega_2=4\pi/\tau$.  As $\tau$ is increased even further this process repeats. This leads to a discontinuous behaviour of the dominant frequency as a function of $\tau$, as shown in Fig. \ref{fig:omegamax}. We stress that the peak locations are only approximately given by $\omega_n=2n\pi/\tau$, careful inspection of the data shown in Fig. \ref{fig:omegamax} reveals quantitative deviations.

The behaviour we have described indicates that delay can have two distinct effects: for $\tau \lesssim 1.2$ the delay dampens the ability of the system to oscillate, whereas for $\tau \gtrsim 1.2$ the delay induces new oscillations in the system. Their characteristic frequency shows a relatively complex dependence on the delay period.

\subsubsection{Spiking behaviour} \label{sec:bruss_spike}

\begin{figure}[t!!]
\centerline{\includegraphics[width=0.5\textwidth]{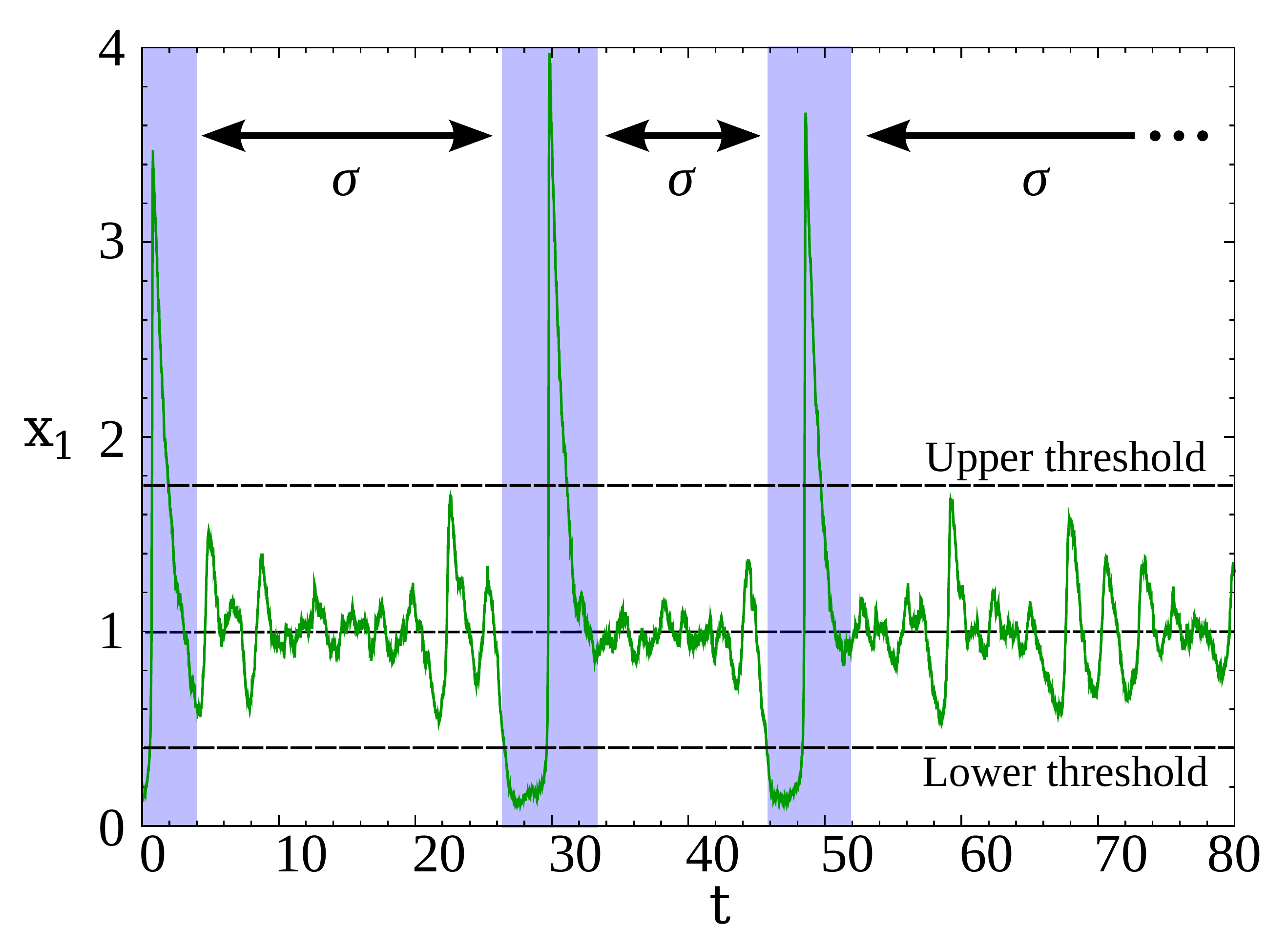}}
\vspace{0.5em}
\caption{(Colour on-line) Example trajectory of $x_1$ for the delay Brusselator with excitable dynamics. Shaded regions correspond to spikes. The Brusselator parameters, $a = 1$, $b = 9.9$, and $c = 9$, are the same as in \cite{giver}. The system size is $\Omega = 100$. The delay kernel is $K(t) = \delta(t-\tau)$ with $\tau = 0.01$.}
\label{fig:bruss_switchtrajectory}
\end{figure}

We now focus on a different parameter regime, one in which the Brusselator system with delay becomes an excitable system, and where it shows spiking behaviour \cite{giver}. In this situation the deterministic system has a stable fixed point, and separated from it in phase space a deterministic `excursion' orbit, taking the system along a looped trajectory with significant amplitude and then back to the fixed point. These excursions are triggered from starting points separated from the deterministic fixed point by a finite distance. In the face of noise (either intrinsic or extrinsic) excitable systems typically fluctuate about the deterministic fixed point, and if fluctuations take it across a threshold triggering an excursion, a spike will occur. Once the spike is complete the system spends time near the fixed point again, until a new excursion 
is triggered by noise. To demonstrate this phenomenon in the delay Brusselator system we show a sample trajectory in Fig.~\ref{fig:bruss_switchtrajectory}. The system mostly fluctuates with small amplitude about the a deterministic fixed point, but intermittent spikes are seen as well. This is a phenomenon seen in many models of nonlinear dynamics with and without delay, including e.g. models of plankton bloom \cite{huppert} or the celebrated FitzHugh-Nagumo model of firing nervous cells \cite{fitzhugh}. 

We will now use the CLE approach to study this phenomenon in more detail. We denote the time between spikes by $\sigma$, and measure the average time between spikes in simulations. Some care needs to be taken here to identify spikes in simulations. We identify the beginning of a spike as a point in time in which the concentration $x_1$ crosses a lower threshold from above, see Fig. \ref{fig:bruss_switchtrajectory}, and the spike is taken to end when the trajectory of $x_1$ has not crossed either the upper or lower thresholds for a fixed time, $t_s=2$. In the example shown in Fig. \ref{fig:bruss_switchtrajectory} the deterministic fixed point corresponds to $x_1^*=1$, the upper and lower thresholds are chosen as $x_l=0.4$ and $x_u=1.75$. The location of the two thresholds and the value of $t_s$ are in principle arbitrary, however the choices we make result in a satisfactory identification of spikes.

\begin{figure}[t!!]
\centerline{\includegraphics[angle =270,width=0.5\textwidth]{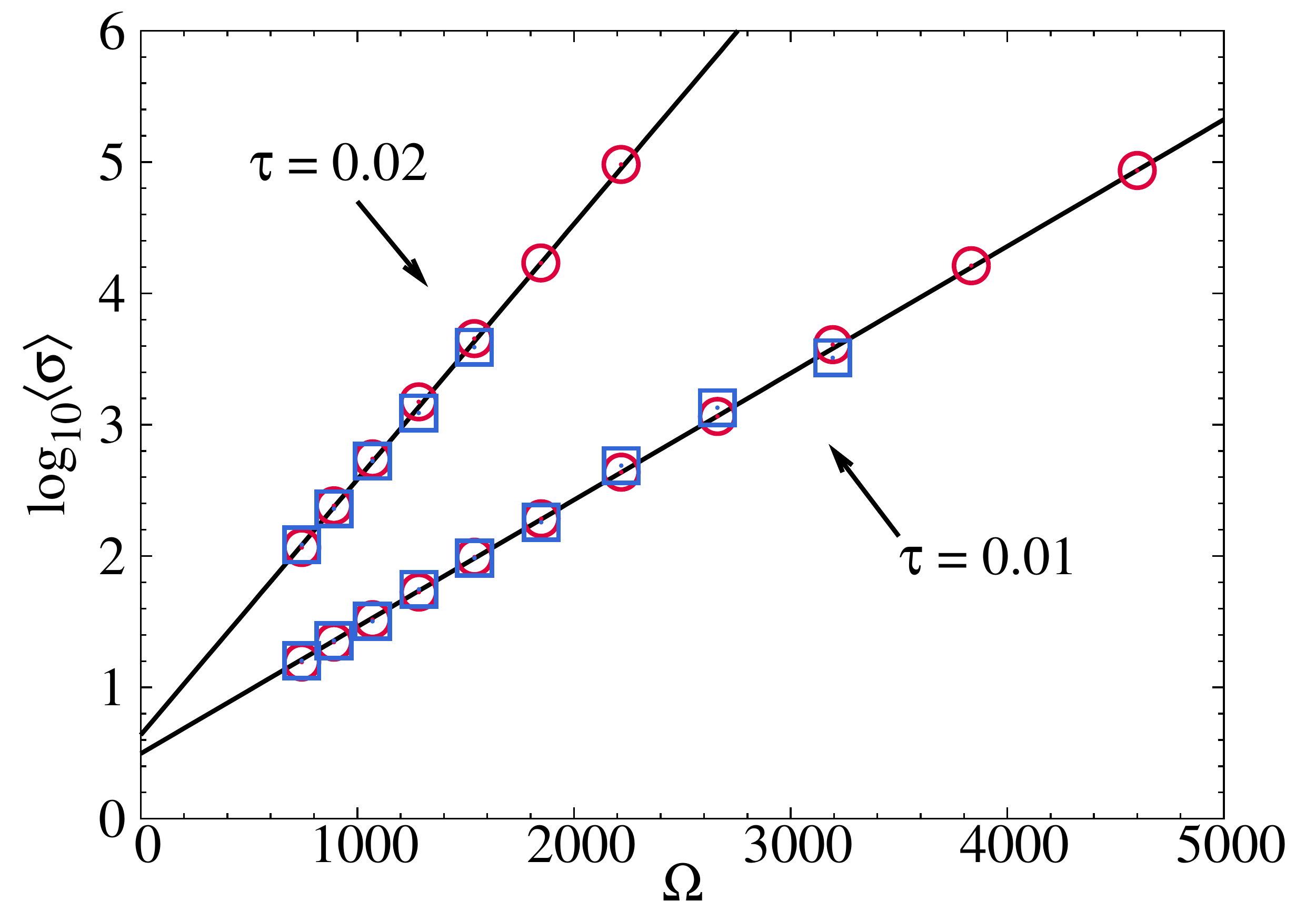}}
\vspace{0.5em}
\caption{(Colour on-line) Mean time between spikes, $\avg{\sigma}$, as a function of $\Omega$ for the delay Brusselator. Blue squares correspond to MNRM simulations, ran up to a final time $t_f = 81920$. Red circles correspond to simulations of the CLE, run up to time $10t_f$ with a time step $\Delta t = 10^{-5}$. The black line shows a fit of an exponential function to the data from the MNRM simulations, with which the CLE simulations can be seen to agree. Parameters used are the same as Fig.~\ref{fig:bruss_switchtrajectory}, apart from $\tau$ and $\Omega$ which are as indicated.}
\label{fig:bruss_switchscaling}
\end{figure}

Fig.~\ref{fig:bruss_switchscaling} shows the mean time between spikes, $\avg{\sigma}$, in the delay Brusselator. Exact simulations using the MNRM are in very good agreement with simulations of the CLE. As seen in the figure the typical time between spikes increases exponentially with $\Omega$. In the regime of large $\Omega$ efficient simulations of the CLE can be advantageous. As discussed further in Appendix \ref{sec:simmethods} the computing time required to perform simulations of the CLE is largely independent of the system size, whereas computational resources to run the MNRM increase quickly with $\Omega$. The exponential behaviour of the time between spikes is in-line with observations that escape rates from locally stable fixed points in dynamical systems subject to noise scale exponentially with decreasing noise strength \cite{kramersesc}, and the fact that spikes are triggered by excursions of the dynamics from the fixed point crossing a specific threshold \cite{fitzhugh}. In the deterministic limit, $\Omega\to\infty$ the time between spikes is infinite, there are no spikes in the deterministic system. At finite noise strength, the duration and amplitude of any given spike remain roughly constant as $\Omega$ is varied, the spikes are different from the stochastic quasi-cycles discussed above. Quasi-cycles are sustained by noise, if the stochasticity were to be removed or 
reduced in the stationary state the quasi-cycles would disappear (or their amplitude be reduced). The spikes in Fig. \ref{fig:bruss_switchtrajectory} are triggered by noise, but once a spike is triggered it completes irrespective of the stochasticity, fundamentally following a deterministic orbit. The noise does not affect the spikes' amplitude, but only the frequency with which they occur. The time between spiking phases is very sensitive to the delay $\tau$, as seen in Fig.~\ref{fig:bruss_switchscaling}, where we show results for two values of $\tau$. An increase in the delay leads to significantly larger average inter-spike time $\avg{\sigma}$.

It is important to stress that the spiking behaviour results from the combination of intrinsic noise and the nonlinear dynamics of the delay Brusselator system. Neither the deterministic limit, nor the LNA capture the spiking behaviour completely. The spiking orbits are present in the deterministic flow, but the frequency with which they occur cannot be obtained from a deterministic analysis alone. A linear analysis about the fixed point on the other hand neglects all nonlinearity, and hence the spiking orbits are not captured. A recent paper has investigated how the nonlinearity of the CLE can be studied analytically (for systems without delay), and this may provide a starting point for further analytical study of similar effects in delay systems \cite{grima.2013}.

\section{Conclusions}\label{sec:concl}
In summary we have presented an intuitive heuristic derivation for the Gaussian approximation of discrete-particle dynamics with distributed delay. These approximations build on our earlier work \cite{brett_galla}, in which we have used a more formal approach based on functional integrals. Our main result is a chemical Langevin equation for systems with general delay distributions, ready to be applied for the efficient simulation of stochastic delay systems of a large, but finite size. The chemical Langevin equation can be reduced in a linear-noise approximation, which allows one to make further analytical progress, in particular with a view towards calculating stationary probability distributions and correlation properties of the stationary state. 

We have applied our results to the example of a Brusselator system with delay. Comparison against simulations shows that the linear-noise approximation works well. We have characterised the spectra of noise-induced quasi-cycles, and we have shown how simulations of the chemical Langevin equation allow for an efficient computational characterisation of noise-triggered spiking behaviour for parameter choices in which the delay Brusselator becomes an excitable system.

We expect that the procedure for the systematic derivation of Gaussian approximations in systems with distributed delay will be of interest to a number of applications, for example in other chemical reaction systems or in the biological sciences. The approach we have presented here is intuitive and relatively easy to apply and to generalise. Model features that have not been captured so far are the uncertain completion of delay reactions or cases in which a delay reaction can result in several later outcomes, possibly depending on the state of the system at the scheduled completion time. Work is in progress to extend the formalism to such cases \cite{brett_galla_inprep}.

\acknowledgements{T.B. thanks the Engineering and Physical Sciences Research Council (EPSRC) for support. Both authors acknowledge helpful discussions with L. F. Lafuerza.}
\appendix
\section{Continuous-time and discrete-time formulation of CLEs with delay}
\label{appendix:contlimit}

In this appendix we discuss the continuous-time limit of Eqs. (\ref{eq:discretelangevin}, \ref{eq:cledcorr}). The second term on the RHS of Eq. (\ref{eq:discretelangevin}) is just a straightforward discretisation of a deterministic drift term, and requires no further discussion. The coefficients $v_{i,\alpha}$ and $w_{i\alpha}$ in the noise term are not material for taking the continuous-time limit, and neither is the pre-factor $\Omega^{-1/2}$, nor the rate $r_i(\bx_t)$. We therefore focus on an equation of the form
\be
x_{t+\Delta} = x_{t} + \sqrt{\Delta} \sum_{i}\sum_{\tau\geq\Delta}\left( \zeta_{i,t}^\tau + \zeta_{i,t-\tau}^\tau\right),
\label{eq:start2}
\ee
with
\be
\avg{\zeta_{i,t}^\tau\zeta_{j,t'}^{\tau'}}=\delta_{i,j}\delta_{t,t'}\delta_{\tau,\tau'}K_i(\tau)\Delta.\label{eq:dcorrapp}
\ee
Our aim is to justify that this is an appropriate discretization of the continuous-time limit 
\be\label{eq:applang}
\dot{x}(t) = \sum_i \int_0^\infty d\tau \left( \zeta_i(t,\tau)+\zeta_i(t-\tau,\tau)\right),
\ee
with
\be \label{eq:appcorr}
\avg{\zeta_i(t,\tau)\zeta_j(t',\tau')} = \delta_{i,j} \delta(t-t')\delta(\tau-\tau')K_i(\tau).
\ee
Our focus is on checking that the scaling with $\Delta$ is appropriate in Eq. (\ref{eq:start2}) and (\ref{eq:dcorrapp}).

\medskip
In order to verify the above claim we compute the second moment of $x(t)$ at a fixed time $t$ both for the discrete-time dynamics, and for the continuous-time variant.
\\

\underline{Continuous time:}\\
Assuming a zero initial condition, and integrating Eq.~\eqref{eq:applang} with respect to time gives
\be
x(t) = \int_0^t dt'\int_0^\infty d\tau \sum_i [ \zeta_i(t',\tau)+\zeta_i(t'-\tau,\tau)].
\ee
From this it is straightforward to calculate the second moment of $x(t)$, 
\begin{align}
&\avg{x(t)^2} = \sum_{i,j}\int_0^t\int_0^t\int_0^\infty\int_0^\infty dsds'd\tau d\tau'\nonumber \\
&\times\bigg\{ \avg{\zeta_i(s,\tau)\zeta_j(s',\tau')} + \avg{\zeta_i(s,\tau)\zeta_j(s'-\tau',\tau')} \nonumber \\
&+ \avg{\zeta_i(s-\tau,\tau)\zeta_j(s',\tau')} + \avg{\zeta_i(s-\tau,\tau)\zeta_j(s'-\tau',\tau')}\bigg\}.
\end{align}
 
Using Eq. (\ref{eq:appcorr}) and the fact that $\int_0^\infty d\tau~ K_i(\tau)=1$ this leads to
\be
\avg{x(t)^2} = \sum_i\left[ 2t + \int_0^t\int_0^t dsds' \Big\{K_i(s'-s) + K_i(s-s')\Big\}\right].
\label{eq:app2ndmomentx}
\ee
\medskip
\underline{Discrete-time dynamics:}\\
Now consider Eq.~\eqref{eq:start2}. Again assuming a zero initial condition, an iterative application of Eq.~\eqref{eq:start2} gives the following expression for $x_t$, 
\be
x_t = \sqrt{\Delta} \sum_i\sum_{\tau\geq\Delta}\sum_{t'<t} \left[\zeta_{i,t'}^{\tau} + \zeta_{i,t'-\tau}^{\tau}\right].
\ee
From this one has
\begin{align}
\avg{x_t^2} =&~ \Delta\sum_{i,j}\sum_{\tau,\tau'\geq\Delta} \sum_{t',t''<t}\bigg\{ \avg{\zeta_{i,t'}^{\tau}\zeta_{j,t''}^{\tau'}} \nonumber \\
&~~~~~~+ \avg{\zeta_{i,t'}^{\tau}\zeta_{j,t''-\tau'}^{\tau'}} + \avg{\zeta_{i,t'-\tau}^{\tau}\zeta_{j,t''}^{\tau'}} \nonumber \\
&~~~~~~+ \avg{\zeta_{i,t'-\tau}^{\tau}\zeta_{j,t''-\tau'}^{\tau'}} \bigg\}.
\label{eq:appdiscrete2ndmoment}
\end{align}

We now apply Eq. (\ref{eq:dcorrapp}). Using $\Delta\sum_{\tau\geq\Delta}K_i(\tau)=1$ one finds
\begin{align}
\avg{x_t^2} =& \Delta^{2}\sum_i\sum_{\tau\geq 0}\sum_{t',t''<t}K_i(\tau)\Big\{\delta_{t',t''}+ 
\delta_{t',t''-\tau} \nonumber \\
&~~~~~~~~~~~~~+ \delta_{t'-\tau,t''}+ \delta_{t',t''} \Big\} \nonumber \\
=& \sum_i\left[2t + \Delta^{2}\sum_{t',t''<t}\left\{K_i(t''-t')+K_i(t'-t'')\right\} \right]
\end{align}
which reduces to Eq. (\ref{eq:app2ndmomentx}) in the limit $\Delta\to 0$.

\section{Simulation methods}
\label{sec:simmethods}

\subsection{Modified next reaction method}
The numerical simulation of stochastic processes with distributed delay is possible using a variation of the modified next-reaction method (MNRM) algorithm \cite{anderson}.  For each delay reaction an ordered list $s_i$ of scheduled event times has to be set up. We will write $s_i(k)$ for the $k$-th element of the list $s_i$ in the following ($k=1,2,\dots$). The lists will be ordered such that $s_i(k)<s_i(k')$ for $k<k'$. The algorithm proceeds as follows:

\begin{enumerate}
\item[1.] Set initial $x_\alpha$ for all species $\alpha$ and set $t=0$. Set $P_i=0$ and $T_i=0$. For each reaction channel involving delay, initialise the list $s_i$ to contain only one element, $s_i=\{\infty\}$.
\item[2.]  Calculate the reaction rates $r_i(\bx)$, for each reaction.
\item[3.] Generate an independent random number $\rho_i$ for all $i$, each drawn from a uniform distribution over $(0,1]$. Set $P_i=-\ln(\rho_i)$.
\item[4.] For each $i$ set $\Delta t_i =(P_i - T_i)/(\Omega r_i(\bx))$.
\label{item:return}
\item[5.] Set $\Delta =\underset{i}{\min}\{\Delta t_i, s_i(1) - t\}$, where $s_i(1)$ is the first (i.e. lowest) entry in $s_i$. We write $i_0$ for the corresponding reaction, $i_0=\underset{i}{\mbox{arg min}}\{\Delta t_i, s_i(1) - t\}$. If the minimum is attained at $\Delta t_{i_0}$ the selected event corresponds to the initiation of a new reaction. If the minimum is at $s_{i_0}(t)-t$, then the selected event is the completion of a delay reaction.
\item[6.] Increment time by $\Delta$, $t\leftarrow t+\Delta$.
\item[7a.] If the event selected in 5. corresponds to the completion of a delay reaction then
\begin{itemize}
 \item  Update the state of the system: $x_\alpha \leftarrow x_\alpha+ w_{i_0,\alpha}/\Omega$ for all $\alpha$.
\item Delete the first entry in $s_{i_0}$.
\end{itemize}
\item[7b.] If the event selected in 5. corresponds to the initiation of a new reaction
\begin{itemize}
 \item Update the state of the system  $x_\alpha \leftarrow x_\alpha + v_{i_{0},\alpha}/\Omega$ for all $\alpha$. 
\item If reaction $i_{0}$ is a delay reaction:
\begin{itemize}
  \item Draw a random number $\tau$ from $K_{i_0}(\cdot)$.
  \item Add the entry $t+\tau$ to the list $s_{i_0}$.
  \item Sort $s_{i_0}$ so that the list remains in ascending order.
  \end{itemize}
\end{itemize}
\item [8.] Draw a uniform random number $\rho$ from the interval $(0,1]$. Update $P_{i_0} \leftarrow P_{i_0} -\ln(\rho)$.
\item[9.] For each $i$ update $T_i\leftarrow T_i+\Omega r_i(\bx)\Delta$.
\item [10.] Recalculate the reaction rates $r_i(\bx)$.
\item [11.] Go to step 4, or exit if final time of the simulation is reached.
\end{enumerate}

This is a minor change to algorithm 7 presented in \cite{anderson}. The only difference relates to the maintenance of ordered lists for each reaction channel. For systems with fixed delay, as discussed in \cite{anderson}, updating these lists is straightforward, events are completed in the order they are initiated (they all have the same delay period). Extending the algorithm to cover distributed delay requires two steps: (i) drawing a random delay time from the appropriate delay kernel when a delay reaction is initiated, and (ii) sorting the list whenever a a new entry is added, to ensure the entries are always in ascending order.  

For systems without delay the time between events scales as $\Omega^{-1}$ as the reaction rates are all proportional to $\Omega$. The computing time to run the MNRM algorithm up to a fixed total time thus grows linearly in the system size $\Omega$. For models with distributed delay the simulation time increases even faster with $\Omega$ as additional time is needed for the sorting of the list of delay events, the typical length of which itself increases with $\Omega$.

\subsection{Simulation of the chemical Langevin equation}

The alternative approach is to simulate the CLE, Eq. (\ref{eq:cle}). These equations are not an exact representation of the original microscopic model, they are an approximation for large, but finite systems. As such, simulations of the CLE will not be exact. On the other hand, simulation times of the CLE are independent of the system size, $\Omega$. This scale only enters through the noise amplitude, and it does not affect the run time needed to integrate the CLE numerically. For large $\Omega$ the simulation of the CLE is not only accurate (in this limit it represents the original process faithfully), but also faster than simulation of the discrete-particle model. In practice the discretisation of the CLE is implemented using Eqs. (\ref{eq:discretelangevin}) and (\ref{eq:cledcorr}). In the case of Markovian systems any one Gaussian random variable only enters at one single time, see Eq. (\ref{eq:differenceeq2}), and they can hence be removed from the computer memory once they have been 
used. For delay systems it is necessary to store the $\zeta_{i,t}^\tau$, generated at time $t$, for later use at $t+\tau$.


\begin{thebibliography}{99}
\bibitem{remark} In Eq. (5) of our earlier paper \cite{brett_galla} we reported a similar expression before the LNA was made, i.e. with $\bx$ instead of $\bx^\infty$ as an argument of $B_{\alpha\beta}$. The precise interpretation of the average $\avg{\dots}$  in that equation is ambiguous, which is why we use a compact noise correlator $B_{\alpha\beta}$ only when the LNA is made in the present paper. The appropriate interpretation of the notation in our earlier paper is given by Eqs. (\ref{eq:cle})-(\ref{eq:zetacorr}) of the present manuscript.
\bibitem{footnote1} The convention $\widetilde f(\omega) = \int_{-\infty}^{\infty} dt e^{-i\omega t}f(t)$ for the Fourier transform is used throughout.
\end{thebibliography}
\end{document}